\documentclass[journal,comsoc]{IEEEtran}
\usepackage{setspace}
\usepackage{cite}
\ifCLASSINFOpdf
\usepackage[pdftex]{graphicx}
\else
\usepackage[dvips]{graphicx}
\fi
\usepackage{amsmath,amssymb,amsthm}

\usepackage{caption}
\usepackage{amsfonts}
\usepackage{algorithm,algorithmic}
\usepackage{array}
\usepackage{makecell}				
\usepackage{cite}
\usepackage{esint}
\usepackage{enumerate}
\usepackage{times}
\usepackage{url}
\usepackage{stfloats}
\usepackage{multirow}
\usepackage{subfigure}
\usepackage{booktabs}
\ifCLASSOPTIONcompsoc
\usepackage[caption=false,font=normalsize,labelfont=sf,textfont=sf]{subfig}
\else
\usepackage[caption=false,font=footnotesize]{subfig}
\fi

\newcommand{\PreserveBackslash}[1]{\let\temp=\\#1\let\\=\temp}
\newcolumntype{C}[1]{>{\PreserveBackslash\centering}p{#1}}
\newcolumntype{R}[1]{>{\PreserveBackslash\raggedleft}p{#1}}
\newcolumntype{L}[1]{>{\PreserveBackslash\raggedright}p{#1}}

\begin{document}

\title{Joint Active and Passive Beamforming with Sensing-Assisted Discrete Phase Shifts for Dual-RIS ISAC Systems}

\author{Qing Xue, \textit{Senior Member, IEEE}, Yun Lan, Jiajia Guo, \textit{Member, IEEE}, Qianbin Chen, \textit{Senior Member, IEEE}, and Shaodan Ma, \textit{Senior Member, IEEE}
\thanks{Qing Xue, Yun Lan and Qianbin Chen are with the School of Communications and Information Engineering, Chongqing University of Posts and Telecommunications, Chongqing 400065, China. (e-mail: xueq@cqupt.edu.cn; s230101070@stu.cqupt.edu.cn; cqb@cqupt.edu.cn).}
\thanks{Jiajia Guo is with the National Mobile Communications Research Laboratory, Southeast University, Nanjing 210096, China. (e-mail: jiajiaguo@seu.edu.cn).}
\thanks{Shaodan Ma is with the State Key Laboratory of Internet of Things for Smart City and the Department of Electrical and Computer Engineering, University of Macau, Macau, SAR, China (e-mail: shaodanma@um.edu.mo).}}

\maketitle

\begin{abstract}
  Targeting the requirements of 6G, this paper investigates a semi-passive dual-reconfigurable intelligent surface (RIS)-assisted integrated sensing and communication (ISAC) system, tackling the max-min user signal-to-interference-plus-noise ratio (SINR) problem via joint active and passive beamforming to enhance system performance and ensure user fairness. Addressing this challenge, we first utilize dual RISs for user angle estimation to simplify the solution process of the formulated problem, an efficient alternating optimization algorithm is then developed. Specifically, semi-definite relaxation and the bisection method are employed to solve the transmit beamforming optimization subproblem. For the RIS discrete phase shifts, a sensing-assisted approach is adopted to constrain the optimization search space, with two distinct low-complexity search strategies introduced for different RIS sizes. Numerical simulation results demonstrate that the proposed algorithm achieves performance close to the ideal continuous phase shift benchmark, outperforms conventional discrete phase shift optimization algorithms, and exhibits a significant improvement over single-RIS systems.
\end{abstract}


\section{Introduction}

\IEEEPARstart{I}{ntegrated} sensing and communication (ISAC) is a key technology for 6G, recognized as a critical enabler for numerous emerging applications such as autonomous driving and smart homes\cite{9737357}\cite{9585321}. ISAC achieves functional unification by consolidating sensing and communication capabilities within base stations (BSs) through shared utilization of hardware, signal processing chains, spectral resources, and network infrastructure. This integrated framework enables simultaneous data transmission and environmental information acquisition, ultimately realizing the synergistic convergence of communication and sensing functionalities\cite{7746569,9705498,8999605}. This stems from the growing convergence between communication and sensing in their technological evolution, where higher frequency bands such as millimeter wave (mmWave) and terahertz waves\cite{10422712}\cite{11060909}, larger antenna arrays, and device miniaturization facilitate hardware and spectrum sharing, thereby providing ubiquitous communication and sensing services for emerging technologies.

While ISAC technology undoubtedly holds significant application potential, in complex environments, obstacles often severely degrade the sensing performance of high-frequency band signals, thereby affecting both communication and sensing capabilities. The introduction of Reconfigurable Intelligent Surface (RIS) \cite{9140329} offers an effective solution to this challenge. A traditional RIS typically consists of a planar surface comprising a large number of passive reflecting elements, each capable of independent control over the amplitude and/or phase of incident signals. By directing these reflected signals, it facilitates propagation paths that bypass obstacles, thereby enhancing both signal coverage and reliability. In ISAC systems, sensing performance typically relies on the line-of-sight (LoS) link between the ISAC transmitter and the target. RIS can be deployed to establish a two-hop virtual LoS link for targets within non-LoS (NLoS) coverage \cite{9326394}. This supplementary link can subsequently be exploited to improve both sensing and communication capabilities. Furthermore, considering applications in complex propagation environments, deploying a single RIS may provide limited improvement for ISAC performance. Coordinating multiple RISs has emerged as an effective solution to this challenge. 

In recent years, studies have increasingly focused on RIS-assisted ISAC systems under various configurations and scenarios. For instance, in \cite{B15}, the authors proposed an Alternating Direction Method of Multipliers (ADMM)-based RIS design to optimize the phase shifts. Similarly, the authors of \cite{B3} proposed the problem to maximize the communication rate under constraints on the sensing signal-to-noise ratio (SNR) and the Cram{\'e}r-Rao bound (CRB), respectively. An alternating iterative update algorithm is developed to solve these two problems. The work \cite{B8} proposed a novel joint active and passive beamforming design for RIS-based ISAC systems. The SNR at the user is maximized under a minimum detection probability constraint. In work \cite{S1}, the minimization of the sensing CRB is formulated as the optimization objective, subject to constraints on the user signal-to-interference-plus-noise-ratio (SINR). The problem is solved by transforming the CRB into a tractable modified fisher information matrix. In \cite{C1}, the authors maximize the radar SINR under quality-of-service (QoS) constraints for communication users. The scaling order of the radar SINR with a large number of reflecting elements is derived. The authors of \cite{B12} proposed the problem of maximizing the weighted radar mutual information and the communication rate, achieving improved communication and sensing performance. The aforementioned studies have conducted in-depth theoretical analyses of ISAC optimization problems from multiple perspectives, while research on RIS has also been progressively deepening. Building on various optimization frameworks, studies such as \cite{C1} and \cite{C2} replace conventional RIS with active RIS to amplify signals and mitigate multiplicative fading caused by multi-hop propagation. In \cite{C2}, the authors leverage active RIS to enhance both radar echo signal quality and communication performance, demonstrating its potential for improving angle estimation accuracy. In \cite{S1}, a simultaneously transmitting and reflecting-RIS (STAR-RIS) is proposed. Additionally, sensors are integrated onto the RIS surface to address sensing issues such as significant path loss and clutter interference. The work in \cite{S5A} combines active RIS with STAR-RIS to extend the sensing range, improve sensing accuracy, and increase the system's sum communication rate.

Beyond functional variations, research on system architectures has also become increasingly diverse. In \cite{11177504}, movable antennas are employed at the BS instead of traditional fixed antennas. In work \cite{10254508}, the system performance is evaluated under both single-RIS and dual-RIS configurations, with the dual-RIS setup demonstrating superior communication and sensing capabilities. Meanwhile, the work \cite{B6} investigates a multi-RIS scenario, which enables simultaneous backscatter signal detection from multiple internet of things devices and supports target sensing based on these signals. This approach maintains high performance while adapting to more complex environments. Similarly, in \cite{10746496}, the authors consider a system with multiple full-duplex access points that concurrently perform target detection and multi-user uplink communication, further extending the scalability and applicability of ISAC networks. In work \cite{10366288}, the authors consider imperfect channel state information (CSI) in an ISAC system and leverage sensing to obtain more accurate CSI, thereby improving both communication performance and positional awareness, achieving a mutually beneficial outcome. Moreover, the authors also considered the discrete phase shifts of the RIS, an aspect that has received limited attention in the existing literature.

Most existing ISAC studies prioritize optimizing either communication or sensing performance by treating the other as a constraint, or focus on optimizing a weighted sum of both. However, the potential for mutual enhancement between sensing and communication has been largely overlooked. Meanwhile, most studies on RIS-assisted ISAC assume ideal continuous phase shifts, while practical RIS implementations support discrete phase shifts. Such an assumption introduces quantization errors, resulting in performance that fails to meet expectations. Furthermore, most achievements focus on constructing ISAC optimization frameworks for single-RIS-assisted scenarios. While these approaches enhance spatial degrees of freedom to some extent, their dynamic environmental adaptability and anti-blockage robustness in mobile user/target scenarios remain limited.

Motivated by these observations, this paper investigates a downlink dual semi-passive-RIS-assisted ISAC system. The study focuses on the joint optimization of active and passive beamforming, and leverages sensed user angle information to reduce the computational complexity of the solution. The main contributions are summarized as follows:
\begin{itemize}
\item \textbf{}In contrast to conventional architectures, we propose a novel dual-RIS-assisted ISAC framework, wherein the RISs are equipped with sensing elements to receive echo signals from users, thereby obtaining user related angle information, which in turn facilitates the design of communication-oriented discrete RIS phase shifts.
\item \textbf{}To address the challenge of designing the RIS phase shift matrix in multi-user scenarios, we propose a sensing-assisted algorithm. It first calculates the optimal phase shift values for each user individually, and the minimal range of these values defines a new discrete phase shift constraint for the RIS elements. Based on the narrowed feasible search space for discrete phase shifts, the complexity of the proposed sensing-assisted RIS optimization algorithm can be effectively reduced.
\item \textbf{}To ensure user QoS and service fairness, we formulate a max-min SINR optimization problem involving the joint design of BS beamforming vectors and dual-RIS discrete phase shifts. To solve this fairness-oriented problem, an alternating optimization algorithm is proposed based on a decomposition-and-reconstruction technique.
\item \textbf{}The BS transmit beamforming optimization subproblem is transformed into a feasibility problem and solved using SDR combined with the bisection method. To address the challenge of high computational complexity in solving for discrete phase shifts, two low-complexity search algorithms are proposed, leveraging the narrowed feasible set obtained from sensing information: a sensing-assisted global search algorithm and a sensing-assisted one-dimensional search algorithm. The former achieves superior performance, while the latter offers lower computational complexity and broader applicability.
\end{itemize}

The remainder of this paper is structured as follows. Section II introduces the related work. Section III introduces the system model and formulates the optimization problem. Section IV presents an alternating optimization algorithm, which employs two sensing-based low-complexity search methods to solve the phase shift optimization subproblem. Numerical results and corresponding analysis are provided in Section V. Finally, Section VI concludes the paper.

\textit{Notations}: Throughout this work, vectors and matrices are denoted by boldface lower case and boldface upper case, respectively. The superscripts $(\cdot)^T$, $(\cdot)^{H}$ and $(\cdot)^{-1}$ denote the operations of transpose, Hermitian transpose and inverse, respectively. The Euclidean norm, absolute value, Kronecker product are respectively denoted by $\|\cdot\|$, $|\cdot|$ and $\otimes$. $\mathcal{CN}(0,\sigma^2)$ denotes the circularly symmetric complex Gaussian distribution with zero mean and variance $\sigma^2$. $\text{tr}(\cdot)$ represents the matrix trace. $\text{diag}(\cdot)$ denotes a square diagonal matrix with the elements in $(\cdot)$ on its main diagonal. 

\section{Related Work }
Research on ISAC dates back to as early as the last century. Communication information is embedded within a set of radar pulses using Pulse Interval Modulation, where the radar functions as a missile ranging instrument \cite{4337601}. However, the interplay between the two functionalities remained relatively limited thereafter. With the continuous evolution of radar and communication technologies, subsequent efforts were made to integrate them. This ultimately led to the formal proposal of the joint radar-communications concept in 2006 \cite{4338141}. During the 2010s, research demonstrated that the effects of random communication data in orthogonal frequency division multiplexing radar could be straightforwardly compensated for through simple Fourier Transform (FT) and inverse FT operations\cite{5776640}. With their parallel development, the operational frequencies of radar and communication systems have converged. This trend drove the need for spectrum sharing, leading to the allocation of sub-6 GHz radar bands for shared use in the United States. Furthermore, this convergence is not limited to operational frequencies but also extends to channel characteristics and signal processing methodologies \cite{6824752}. This evolution has led to the erosion of the traditional boundary between radar and communication, with their functionalities transcending dedicated infrastructures. The seamless integration of sensing and communication paves the way for and underpins the paradigm of ISAC \cite{9606831}. Owing to its demonstrated significant potential, research in this field has advanced rapidly and has to date yielded remarkable achievements.

With a focus on RIS-assisted ISAC, a review of existing relevant literature is presented below, examining aspects such as research scenarios, system models, and problem formulation. For clarity, we provide a detailed comparison of related works in Table I.

\begin{table*}[htbp]
	\centering
	\caption{Literature Review on RIS-assisted Integrated Sensing and Communication (ISAC)}
	\label{tab:ris_isac_literature}
	\begin{tabular}{|p{3.2cm}|p{2.2cm}|p{1.8cm}|p{2.8cm}|p{2.8cm}|}
		\hline
		\textbf{Ref.} & \textbf{Scenario} & \textbf{RIS Config.} & \textbf{RIS Type} & \textbf{Optimization Focus} \\
		\hline
		\cite{A3,A5,A9} & Conventional ISAC & N/A & N/A & Sensing-centric \\
		\hline
	    \cite{A8,A10} & Conventional ISAC & N/A & N/A & Communication-centric \\
        \hline
        \cite{A4,A2,A6,A7} & Conventional ISAC & N/A & N/A & Joint optimization \\
        \hline
        \cite{B2,B4,B5,S1,C1,B9,B13,S6,S9,S11,M1,BD1,S13A,C5,S15A,C7,B17,S18,BD2} & RIS-assisted ISAC & Single-RIS & Continuous phase shifts & Sensing-centric \\
		\hline
        \cite{B3,B8,B11,S4,S5A,B15,S16,BD3} & RIS-assisted ISAC & Single-RIS & Continuous phase shifts & Communication-centric \\
		\hline
        \cite{S2,S3,B14,C3,S7,S8,S12A,S14,C6,S17A} & RIS-assisted ISAC & Single-RIS & Continuous phase shifts & Joint optimization \\
		\hline
        \cite{B12} & RIS-assisted ISAC & Single-RIS & Discrete phase shifts & Sensing-centric \\
		\hline
        \cite{B16} & RIS-assisted ISAC & Single-RIS & Discrete phase shifts & Joint optimization \\
		\hline
		\cite{B10} & RIS-assisted ISAC & Multi-RIS & Continuous phase shifts & Sensing-centric \\
		\hline
		\cite{B6} & RIS-assisted ISAC & Multi-RIS & Continuous phase shifts & Communication-centric \\
		\hline
		\cite{B7,B20} & RIS-assisted ISAC & Multi-RIS & Continuous phase shifts & Joint optimization \\
		\hline
		\cite{B21} & RIS-assisted ISAC & Multi-RIS & Discrete phase shifts & Communication-centric \\
		\hline
		\cite{B1,D1,C4,S10} & RIS-assisted ISAC & Dual-RIS & Continuous phase shifts & Communication-centric \\
		\hline
		\cite{B19} & RIS-assisted ISAC & Dual-RIS & Discrete phase shifts & Joint optimization \\
		\hline
	\end{tabular}
\end{table*}
\subsection{Evolution of Research Scenarios}
With the deepening of research, ISAC architectures have become increasingly diversified and sophisticated. The evolution from traditional ISAC \cite{A6, A3, A4, A8, A10} to RIS-assisted ISAC \cite{B2, B3, B4, B5, B8} marks a significant advancement, where the introduction of RIS provides a revolutionary means to address coverage blind spots and enhance spatial degrees of freedom.

In the early stages of ISAC development, research primarily focused on the coexistence of communication and sensing functionalities. For instance, in \cite{A6}, a dual-functional radar-communication system was investigated, where a BS equipped with both communication and radar capabilities simultaneously communicates with downlink users and detects signals in specific azimuth angles. The problem was formulated as a weighted sum of sensing and communication rates, ensuring overall system performance. In \cite{A3} and \cite{A10}, the communication rate was treated as a constraint while maximizing the radar beam pattern power. Meanwhile, the work \cite{A4} proposed a power optimization strategy that minimized transmit power under both sensing and communication constraints, jointly considering the two functionalities. Research scenarios have also diversified. For example, the authors in \cite{A2} integrated ISAC with vehicular communication to provide enhanced communication and sensing services. The work \cite{A5} combined orthogonal time frequency space and spatial spreading with ISAC to achieve reliable communication and high-precision sensing.

As ISAC research advances, it is expected to operate in complex and dynamic environments, whereby introducing RIS serves as a promising approach. The integration of RIS and ISAC has demonstrated considerable potential. For instance, in \cite{B2}, non-orthogonal multiple access (NOMA) was incorporated into a RIS-assisted ISAC system, significantly improving sensing performance. In \cite{B3} and \cite{B9}, the authors derived the CRB for RIS-assisted ISAC systems. The work in \cite{B8} optimized the RIS-assisted ISAC system through beamforming, leading to notable performance gains. This has spurred numerous subsequent studies on beamforming optimization\cite{B4, B11, B12, B15, B17}. Traditional ISAC systems are highly susceptible to environmental variations, which often lead to performance degradation. The incorporation of RIS transforms ISAC from passively adapting to the environment to actively shaping it, substantially enhancing system performance and robustness.

Under more complex scenarios involving mobility or clutter-rich environments, the performance enhancement of single-RIS-assisted ISAC systems may fall short of expectations. Recent research has evolved from single-RIS architectures\cite{B3, B4, B8} toward multi-/dual-RIS configurations \cite{B1, D1, B6, B7, B10, B20}, leveraging inter-RIS coordination for enhanced performance. Early studies on RIS-assisted ISAC primarily focused on multi-user and multi-target scenarios\cite{B3, B4, B13}, where a single-RIS demonstrated notable performance improvements. However, physical limitations have pushed such configurations to their bottleneck, prompting investigations into multi-/dual-RIS deployments to overcome these constraints. In \cite{B1}, a dual-RIS assisted ISAC system was proposed, jointly optimizing the beamforming of both RIS and radar to mitigate mutual interference. The work \cite{D1} assigned distinct functionalities to two RISs, one dedicated to sensing and the other to communication, thereby enabling simultaneous execution of both services. The work \cite{B6} derived the CRB for a multi-RIS assisted ISAC system, maximizing the total system rate under sensing constraints. In \cite{B7} and \cite{B20}, a novel multi-RIS assisted ISAC structure was introduced, utilizing two RISs for sensing while maintaining simultaneous communication, achieving high-precision positioning and communication performance comparable to communication-only scenarios. The work \cite{B10} proposed a distributed RIS framework for indoor wireless communication, where coordinated multi-RIS operation mitigated signal blockage caused by moving occupants. Beyond RIS, the utilization of multiple BSs and access points constitutes a promising research direction for RIS-assisted ISAC systems \cite{10411853,10746496}.

\subsection{Model Refinement}

Apart from differences in model architectures, the research focus also varies significantly even within the same architectural framework. Early research on RIS-assisted ISAC systems predominantly assumed continuous phase shifts for the RIS elements \cite{B4, B5, B8, B11, B15} to simplify problem formulation and establish performance upper bounds. For instance, \cite{B4} investigated the joint robust design of the BS transmit beamformer and the RIS phase shifts, formulating a worst-case robust beamforming problem. The work \cite{B5} introduced a novel estimation method that enables multiple measurements for direction-of-arrival (DOA) estimation by controlling the RIS reflection matrix, and derived the theoretical Cram{\'e}r-Rao Lower Bound for vehicle DOA estimation in the ISAC system. In \cite{B11} and \cite{B15}, an ADMM-based RIS design was proposed to address non-convex radar optimization problems, extending the algorithm to optimize RIS phase shifts. As research progresses, discrete RIS phase shift models \cite{B12, B16, B19, B21} have gained increasing attention due to their better alignment with practical hardware constraints, spurring the development of various low-complexity quantization algorithms. However, the finite and non-convex nature of discrete phase shifts significantly increases the difficulty of problem solving. In \cite{B19}, an uplink data transmission problem was considered to simultaneously achieve communication and positioning, proposing two beamforming algorithms for sensing and communication that utilize probability matrices to handle discrete phase shifts. The work \cite{B12} developed an alternating optimization algorithm based on the Riemannian Gradient method to solve the non-convex problem introduced by discrete phase shifts. The work \cite{B16} incorporated practical RIS constraints including unit modulus and discrete phase shifts. The work \cite{B21} employed a Zero-Forcing algorithm to eliminate interference and proposed a successive refinement algorithm to iteratively solve the discrete phase shift problem, obtaining a suboptimal solution.

In addition, the functionality of RIS has evolved from simple passive signal reflection toward active, transmissive, and integrated sensing capabilities. To address signal fading challenges, active RIS architectures have been developed by integrating low-cost amplification components, enabling signal amplification in addition to phase adjustment. In \cite{C1}, an active RIS was introduced to maximize radar SINR under communication rate constraints, with the scaling order of radar SINR derived for large reflecting elements. The work \cite{C2} employed active RIS to overcome limitations caused by weak echo signals and derived the corresponding CRB. The work \cite{C4} proposed a dual active RIS structure to mitigate multiplicative fading, while \cite{C5} demonstrated the performance advantages of active RIS over passive RIS through comparative analysis. The work \cite{C3} investigated a hybrid RIS configuration, showing that even partial integration of active elements can enhance system performance. Furthermore, \cite{C6} introduced an active RIS with antenna selection capability, addressing the high energy consumption of Radio Frequency chains in conventional active RIS and facilitating its large-scale deployment. To further improve coverage, STAR-RIS has been designed, enabling 360 $^\circ$ communication and sensing by concurrently manipulating transmitted and reflected signals. In \cite{S11}, STAR-RIS was applied to multi-target detection and communication scenarios, where joint beamforming design between the dual-functional BS and STAR-RIS effectively suppressed clutter interference and enhanced sensing performance. The work \cite{S3} explored STAR-RIS deployment in high-mobility environments. Given its suitability for multi-target and multi-user scenarios, several studies have integrated non-orthogonal multiple access with STAR-RIS \cite{S2, S4, S6, S9} to mitigate multi-user interference prior to signal decoding. The broad coverage of STAR-RIS also raises security concerns, leading to extensive research on anti-eavesdropping techniques in STAR-RIS-assisted ISAC systems \cite{S4, S8, S15A}. Additionally, \cite{S5A} and \cite{S12A} combined active RIS with STAR-RIS, achieving notable performance gains in multi-target and multi-user scenarios. The work \cite{S13A} proposed a hybrid active-passive STAR-RIS architecture, where passive reflective elements serve communication users while active transmitting elements enhance sensing performance, overcoming high path loss in multi-hop transmissions for distant users and targets outside BS coverage. 

With the continuous advancement of related research, an increasing number of novel RIS architectures are being incorporated into ISAC systems, such as Beyond Diagonal-RIS (BD-RIS) and Multi-functional-RIS (MF-RIS). In works \cite{BD1, BD2, BD3}, BD-RIS-assisted ISAC systems are proposed. By introducing an internal interconnection network among scattering elements that enables energy flow and mutual coupling between units, the response matrix becomes a fully non-diagonal matrix. This enhances beamforming flexibility and system robustness. In \cite{M1}, MF-RIS is utilized in an ISAC system, enabling simultaneous sensing, communication, computation, and power transfer, thereby expanding the functional dimensions of ISAC systems. 

\subsection{Diversification of Optimization Objectives}

The continuous diversification of RIS types has led to increasingly varied functional roles within ISAC systems, gradually forming two dominant paradigms: communication-centric and sensing-centric designs. Research on communication-centric methods \cite{S4, C4, S10, S16} and sensing-centric approaches \cite{S11, S1, S6, S9} reveals the distinct roles of RIS across different application scenarios. In communication-centric studies, the work in \cite{S4} maximizes the communication rate while satisfying the minimum required beam pattern gain for sensing. Similarly, the works \cite{S10} and \cite{S16} focus on maximizing the sum communication rate under the constraint of maintaining specified target detection performance. In sensing-centric research, the authors in \cite{S6} and \cite{S11} maximize the minimum sensing beam pattern gain subject to communication rate constraints. The works in \cite{S1} use the CRB as the sensing performance metric, aiming to minimize the CRB while satisfying communication requirements. In \cite{S9}, the authors formulate a matching error minimization problem to adapt the sensing beam pattern while ensuring a minimum communication rate. Additionally, the work \cite{S18} employs the sensing SNR to evaluate sensing performance. This systematic categorization demonstrates how RIS configurations can be dynamically optimized to address the inherent trade-offs between communication and sensing functionalities in ISAC systems.

To further enhance the system's adaptability in dynamic environments, weighted joint optimization of communication and sensing \cite{A7, S2, S3, S7, S14} has emerged as a mainstream approach, aiming to dynamically balance the performance of both functionalities with enhanced universality. In \cite{S2}, the authors formulated a fairness maximization problem between communication users and sensing targets through joint sensing-communication optimization. The work \cite{S3} proposed a simultaneous transmission framework for STAR-RIS-assisted ISAC. To achieve performance balance between sensing and communication functionalities, the framework optimizes energy splitting factors to realize such equilibrium. Furthermore, balancing sensing and communication performance can also be achieved through optimization of energy efficiency \cite{S14}\cite{S17A} and power allocation \cite{B14}.

\subsection{Summary}

\begin{itemize}
\item \textbf{}The development of ISAC has not been confined to its original paradigm. Its integration with RIS has significantly expanded its capabilities and effectiveness. However, in multi-user and multi-target scenarios, single-RIS configurations face limitations due to constrained spatial degrees of freedom and insufficient beamforming flexibility. In contrast, dual-RIS and multi-RIS architectures, which effectively address these challenges, demonstrate greater potential for future development. 
\item \textbf{}The evolution from idealized continuous RIS phase shifts to practical discrete phase shifts reflects a growing alignment with real-world application requirements. Concurrently, the expanding functionality of RIS offers increasingly diverse solutions to challenges encountered in ISAC systems. This progression demonstrates how cross-technology integration provides effective approaches for addressing complex system-level problems.
\item \textbf{}The focus in ISAC systems has shifted from prioritizing either sensing or communication to jointly optimizing and balancing their performance, leading to increasingly tight integration between the two functions. However, existing approaches remain constrained by a fundamental limitation: they treat sensing and communication as independent functionalities. This ``isolated optimization" inevitably leads to performance trade-offs and compromises, whereas the true potential of ISAC systems lies in the deep integration and synergy between sensing and communication---such as sensing-assisted communication or communication-assisted sensing.
\end{itemize}

\section{System Model}

As shown in Fig. 1, we consider a dual-RIS-assisted mmWave ISAC system, in which the LoS links between the BS and $K$ single-antenna users are blocked. To ensure adequate spatial coverage and beamforming flexibility, two semi-passive RISs are deployed to reconstruct the communication and sensing links. Meanwhile, the BS has a uniform linear array with ${{N_t}}$ antenna elements, ${N_t} > K$. Each semi-passive RIS consists of a rectangular array with $N = {N_y} \times {N_z}$ reflective elements and a rectangular array with $M = {M_y} \times {M_z}$ sensing elements. The BS and RIS are connected via a backhaul link and controlled by an RIS controller to facilitate information exchange.

\begin{figure}[h]
    \centering
    \includegraphics[width=1\linewidth]{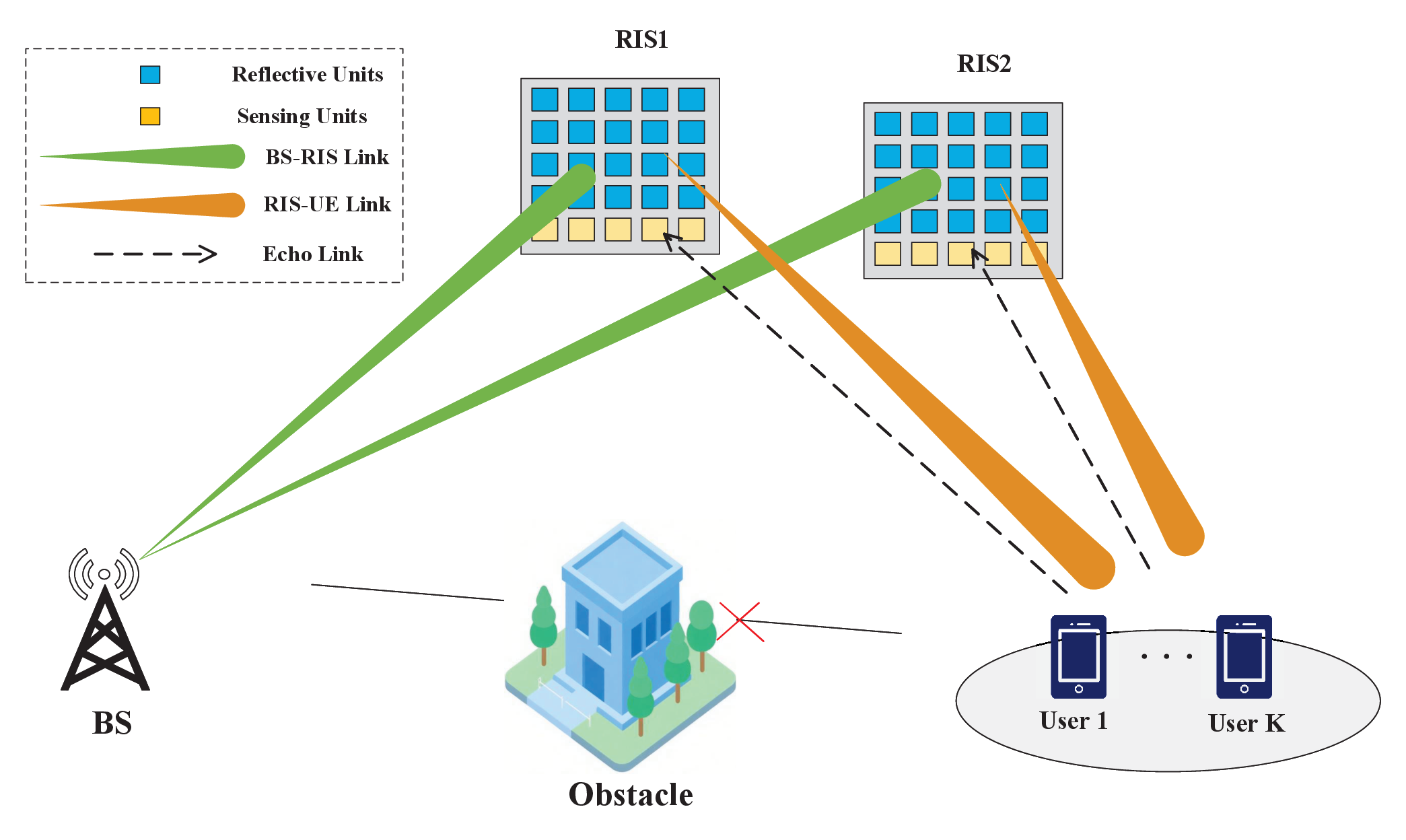}
    \caption{A dual-RIS-assisted ISAC system.}
\end{figure}

 This study divides a transmission cycle into a sensing phase and a communication phase, as show in Fig. 2. The sensing information is utilized to optimize communication performance. Specifically, during the sensing phase, the BS transmits sensing signals, while the RIS randomly generates a phase shift matrix to reflect the signals to the users. Then, the RIS sensing elements receive echo signals from the user and estimates relevant link angle information. To avoid angular interference that may arise from the reflected signals, the two RISs operate alternately during this sensing procedure. Subsequently, the RISs simultaneously assist in the transmission of communication signals. In this phase, we design the phase shift matrix of each RIS based on the estimated angle information acquired in the sensing stage.
\begin{figure}[h]
    \centering
    \includegraphics[width=1\linewidth]{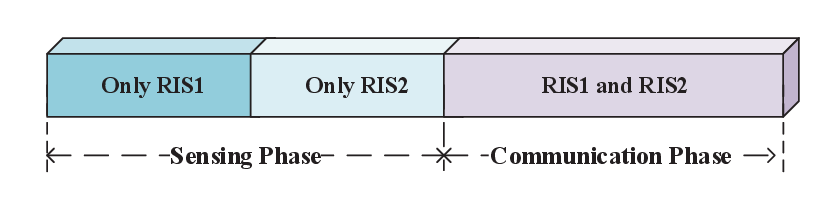}
    \caption{The transmission cycle structure.}
\end{figure}
\subsection{Signal Model}

It is assumed that the ISAC system employs two RISs with identical functionality and the same number of array elements. The phase shift matrix of RIS $i$ is given by ${{\bf{\Theta }}_i} = diag\left( {{{{\theta }}_i}} \right)$ where $\forall i \in\{1,2\}$, with the beamforming vector as  ${{{\theta }}_i} = {\left[ {{\beta _{i,1}}{e^{j{\theta _{i,1}}}}, \cdots ,{\beta _{i,N}}{e^{j{\theta _{i,N}}}}} \right]^T} \in {\mathbb{C} ^{N \times 1}}$, where ${\beta _{i,n}} \in \left[ {0,1} \right]$ is the reflection amplitude of the $n$-th RIS element. To maximize the communication rate, we assume ${\beta _{i,n}} = 1,\forall n \in \left\{ {1, \cdots ,N} \right\},\forall i \in\{1,2\}$. In practical implementations, RISs typically employ discrete phase shifts, meaning each phase shift ${\theta _{i,n}}$ can only take values from uniformly quantized intervals within $\left[ {0,2\pi } \right)$ with uniform quantization intervals. Specifically,
\begin{align}
{\theta_{i,n}}\in  \mathcal{F}\triangleq \left[ {0,\Delta \theta ,...,\left( {{2^b} - 1} \right)\Delta \theta } \right], \forall n \in \left\{ {1, \cdots ,N} \right\},
\end{align}
where $b$ denotes the number of quantization bits, ${2^b}$ is the number of phase-shift levels, and $\Delta \theta  = \frac{{2\pi }}{{{2^b}}}$.

At the beginning of the transmission cycle, ${x_1}\left( t \right)$ is the signal sent by the BS at snapshot $t$, with ${{\bf{w}}_s}\in {\mathbb{C} ^{N_t\times 1}}$ denoting the sensing beamforming vector. The echo signal received at RIS $i$ is expressed as
\begin{align}
{{\bf{y}}_i}\left( t \right) =  {{{\bf{H}}_{{{\mathop{\rm R}\nolimits}}{{\mathop{\rm U}\nolimits}}{{\mathop{\rm S}\nolimits} ,i}}}{{\bf{\Theta }}_i}{{\bf{H}}_{{\mathop{\rm B}\nolimits} {{\mathop{\rm R}\nolimits} ,i}}}{{\bf{w}}_s}{x_1}\left( t \right)}  + {{\bf{n}}_i}\left( t \right),
\end{align}
where ${{{{\bf{H}}_{{\mathop{\rm B}\nolimits} {{\mathop{\rm R}\nolimits} ,i}}}}}\in {\mathbb{C} ^{N \times {N_t}}}$ is the channel matrix from the BS to RIS $i$, ${{{{\bf{H}}_{{{\mathop{\rm R}\nolimits}}{{\mathop{\rm U}\nolimits}}{{\mathop{\rm S}\nolimits} ,i}}}}} \in {\mathbb{C} ^{M \times N}}$ represents the echo channel from RIS $i$ to user, and ${{\bf{n}}_i}\left( t \right) \sim \mathcal{C}\mathcal{N}\left( {0,\sigma ^2{{\bf{I}}_N}} \right)$ is additive white Gaussian noise (AWGN) with covariance matrix $\sigma^2$.

The ISAC system then proceeds to the communication phase, during which the BS transmits composite signal ${\bf{x}} = \sum\nolimits_{k = 1}^K {{{\bf{w}}_k}{s_k}} $, where ${{\bf{w}}_k} \in {\mathbb{C}^{{N_t} \times 1}} $ is the beamforming vector, ${s_k}$ is the data symbol, $\forall k \in \mathcal{K} = \left\{ {1,2,...,K} \right\}$. The received signal at user $k$ is given by
\begin{align}
{{\mathop{\rm y}\nolimits} _k} = \left( {{\bf{h}}_{1,k}^H{{\bf{\Theta }}_1}{{\bf{H}}_{{{{\mathop{\rm BR}\nolimits} },1}}} + {\bf{h}}_{2,k}^H{{\bf{\Theta }}_2}{{\bf{H}}_{{{{\mathop{\rm BR}\nolimits} },2}}}} \right)\sum\nolimits_{k = 1}^K {{{\bf{w}}_k}{s_k}}  + {n_k}, \end{align}
where ${{\bf{h}}_{i,k}} \in {\mathbb{C}^{N \times 1}}$ is the channel vector from the RIS $i$ to user $k$, ${n_k} \sim \mathcal{C}\mathcal{N}\left( {0,{\sigma_k^2}} \right)$ is the AWGN.

Hence, the SINR at user $k$ can be expressed as
\begin{align}
{\rm{SIN}}{{\rm{R}}_k} = \frac{{{{\left| {\left( {{\bf{h}}_{1,k}^H{{\bf{\Theta }}_1}{{\bf{H}}_{{{{\mathop{\rm BR}\nolimits} },1}}} + {\bf{h}}_{2,k}^H{{\bf{\Theta }}_2}{{\bf{H}}_{{{{\mathop{\rm BR}\nolimits} },2}}}} \right){{\bf{w}}_k}} \right|}^2}}}{{\sum\limits_{j = 1,j \ne k}^K {{{\left| {\left( {{\bf{h}}_{1,k}^H{{\bf{\Theta }}_1}{{\bf{H}}_{{{{\mathop{\rm BR}\nolimits} },1}}} + {\bf{h}}_{2,k}^H{{\bf{\Theta }}_2}{{\bf{H}}_{{{{\mathop{\rm BR}\nolimits} },2}}}} \right){{\bf{w}}_j}} \right|}^2} + {\sigma_k^2}} }}.
\label{eq:SINR}
\end{align}

\subsection{Channel Model}

In this study, both the BS-RIS links and RIS-user links are assumed to exhibit LoS propagation. A geometric channel model is employed to characterize the mmWave channels \cite{10366288}
\begin{align}
\mathbf{H}_{\mathrm{BR},i} &= \alpha _{1,i} \mathbf{a}_{\mathrm{R}}\left(u_{\mathrm{BR},i}, v_{\mathrm{BR},i}\right) \mathbf{b}_{\mathrm{B}}^H\left(v_{\mathrm{BR},i}\right), \\
\mathbf{h}_{i,k}^H &= \alpha _{2,i} \mathbf{a}_{\mathrm{R}}\left(u_{i,k}, v_{i,k}\right), \\
\mathbf{H}_{\mathrm{R} \mathrm{U} \mathrm{S},i} &= \sum\nolimits_{k = 1}^K\alpha _{3,i} \mathbf{a}_{\mathrm{S}}\left(u_{k,i}, v_{k,i}\right) \mathbf{a}_{\mathrm{R}}^H\left(u_{i,k}, v_{i,k}\right),
\end{align}
where ${\alpha _{1,i}}$, ${\alpha _{2,i}}$ and ${\alpha _{3,i}}$ denote the complex channel gains of the BS to RIS $i$, RIS $i$ to user $k$, and RIS $i$ receiving echo signals from user $k$, respectively. Meanwhile, we define the angle parameters as follows: the angle of departure (AoD) ${v_{{\rm{B}}{{\rm{R}},i}}}$ at the BS, the angle of arrivals (AoAs) $\left( {{u_{{\rm{B}}{{\rm{R}},i}}},{v_{{\rm{B}}{{\rm{R}},i}}}} \right)$ and AoDs $\left( {{u_{i,k}},{v_{i,k}}} \right)$ at RIS $i$, and the AoAs $\left( {{u_{k,i}},{v_{k,i}}} \right)$ of the user echo at RIS $i$. Moreover, ${\bf{b}}_{\rm{B}}^T\left( {{v_{{\rm{B}}{{\rm{R}},i}}}} \right),{{\bf{a}}_{\rm{R}}}\left( {{u_{{\rm{B}}{{\rm{R}},i}}},{v_{{\rm{B}}{{\rm{R}},i}}}} \right),{{\bf{a}}_{\rm{S}}}\left( {{u_{k,i}},{v_{k,i}}} \right)$ are the array response vectors of the BS, RIS $i$ reflection element, and RIS $i$ sensing element, respectively, which are given by
\begin{align}
\mathbf{b}_{\mathrm{B}}(v) &= \left[ 1, \cdots, e^{j(n-1)v}, \cdots, e^{j(N_t-1)v} \right]^T, \\
\mathbf{a}_{\mathrm{R}}(u,v) &= \left[ 1, \cdots, e^{j(n-1)u}, \cdots, e^{j(N_y-1)u} \right]^T \nonumber \\
&\quad \otimes \left[ 1, \cdots, e^{j(n-1)v}, \cdots, e^{j(N_z-1)v} \right]^T, \\
\mathbf{a}_{\mathrm{S}}(u,v) &= \left[ 1, \cdots, e^{j(n-1)u}, \cdots, e^{j(M_y-1)u} \right]^T \nonumber \\
&\quad \otimes \left[ 1, \cdots, e^{j(n-1)v}, \cdots, e^{j(M_z-1)v} \right]^T.
\end{align}

In addition, we have \cite{9181610}
\begin{align}
\label{eq:steering_vector}
u &= 2\pi \frac{d}{\lambda} \cos(\varphi) \sin(\phi), \\
v &= 2\pi \frac{d}{\lambda} \sin(\phi),
\label{eq:steering_vector2}
\end{align}
where $\varphi $ represents the elevation angle, $\phi $ denotes the azimuth angle, ${d_{{\mathop{\rm RIS}\nolimits} }}$ and ${d_{{\mathop{\rm BS}\nolimits} }}$ represent the distances between two adjacent reflecting elements of the RIS and two adjacent antennas of the BS, respectively. We assume that ${d = d_{{\mathop{\rm RIS}\nolimits} }} = {d_{{\mathop{\rm BS}\nolimits} }} = \frac{\lambda }{2}$, where $\lambda $ denotes the carrier wavelength. In addition, under the far-field condition, the AoAs from the user to RIS sensing element can be assumed to be the same as the AoDs from the RIS reflection elements to user, and will be uniformly denoted as $\left( {{u_{k,i}},{v_{k,i}}} \right)$ hereafter\cite{9724202}.

\subsection{Problem Formulation}

For the dual-RIS-assisted multi-user ISAC system, this paper aims to maximize system performance through joint optimization of the BS transmit beamforming vector ${{\bf{w }}_k}$ and the RIS phase shift matrix ${{\bf{\Theta }}_i}$, while ensuring user service fairness. For RIS-assisted ISAC systems, most prior works on joint beamforming design typically make idealized assumptions about continuous phase shifts at the RIS, or employ simple quantization methods to convert continuous phases into discrete values. Such approaches may suffer from cumulative quantization errors during the conversion process, leading to suboptimal performance in practical implementations. To better align with practical constraints, we directly consider the multi-user ISAC performance optimization problem under discrete phase shift constrains, formulated as
\begin{equation}
\begin{aligned}
& \max_{{{\bf{w}}_k}, \boldsymbol{\Theta}_1, \boldsymbol{\Theta}_2} \min_{k \in \mathcal{K}} \ \mathrm{SINR}_k \\
& \mathrm{s.t.} \quad
\begin{aligned}[t]
& \mathrm{C1}: \mathrm{SINR}_k \geq \gamma ,\\
& \mathrm{C2}: \sum_{k=1}^K \|{{\bf{w}}_k}|^2 \leq P_{\max} ,\\
& \mathrm{C3}: \theta_{i,n} \in \mathcal{F},
\end{aligned}
\end{aligned}
\label{eq:aligned_problem}
\end{equation}
where $\gamma$ represents the minimum SINR requirement for users, and $P_{\max}$ limits the maximum power at the BS. C3 is the discrete phase shift constraint for the RIS. Since the beamforming vector ${{\bf{w}}_k}$ and phase shift matrices $\boldsymbol{\Theta}_i$ in the problem are coupled together, the problem becomes non-convex and is difficult to solve directly.

\section{Problem Solution}

For the multi-user dual-RIS-assisted ISAC system, the optimization objective of the formulated problem~\eqref{eq:aligned_problem}~is to enhance the performance of the worst-case user, thereby ensuring fairness. This section is dedicated to designing sensing-assisted optimization strategies. For solving the challenging problem, we first introduce an auxiliary variable $\tau $ to represent the minimum SINR value across all users, thereby transforming the original problem into a maximization problem

\begin{equation}
\begin{aligned}
& \max_{{{\bf{w}}_k}, \boldsymbol{\Theta}_1, \boldsymbol{\Theta}_2} \tau \\
& \mathrm{s.t.} \quad
\begin{aligned}[t]
& \mathrm{C1}': \mathrm{SINR}_k \geq \tau ,\\
& \mathrm{C2}: \sum_{k=1}^K \|{{\bf{w}}_k}|^2 \leq P_{\max} ,\\
& \mathrm{C3}: \theta_{i,n} \in \mathcal{F}.\\
\end{aligned}
\end{aligned}
\label{eq:problem 2}
\end{equation}

When the max-min problem has a solution, the condition $\tau  \geq \gamma$ must hold, thus, the minimum SINR requirement of the original problem~\eqref{eq:aligned_problem}~is replaced by constraint $\text{C1}'$. In the worst case, we have $\tau = \gamma$.

Subsequently, to resolve the non-convexity caused by coupled variables, the problem~\eqref{eq:problem 2}~is decomposed into three subproblems: the transmit beamforming vector ${{\bf{w}}_k}$ optimization subproblem at the BS and the phase shift matrices $\boldsymbol{\Theta}_1$ and $\boldsymbol{\Theta}_2$ optimization subproblems at the RISs. Here, the former subproblem is addressed via semi-definite relaxation (SDR) technique through its transformation into a feasibility problem, while the latter employs a search-based method that leverages the sensed angle information from the RIS to reduce the computational complexity. An iterative optimization approach is then employed to solve these subproblems alternately.

\subsection{Optimization of BS Beamforming Vector}

To optimize the transmit beamforming vector ${{\bf{w}}_k}$, given the phase shift matrices of $\boldsymbol{\Theta}_1$ and $\boldsymbol{\Theta}_2$, problem~\eqref{eq:problem 2}~is simplified as
\begin{equation}
\begin{aligned}
& \max_{{{\bf{w}}_k}}  \quad \tau \\
& \text{s.t.} \quad
\begin{aligned}[t]
& \text{C1}': \text{SINR}_k \geq \tau,\\
& \text{C2}: \sum_{k=1}^K \|{{\bf{w}}_k}|^2 \leq P_{\max}. \\
\end{aligned}
\end{aligned}
\label{eq:problem 3}
\end{equation}

This optimization problem remains non-convex. The SDR technique can be employed to transform it into a more tractable form for solution. We define the combined channel from the BS to user $k$ as
${{\bf{h}}_k}={\bf{h}}_{1,k}^H{{\bf{\Theta }}_1}{{\bf{H}}_{{\rm{B}}{{\rm{R}},1}}} + {\bf{h}}_{2,k}^H{{\bf{\Theta }}_2}{{\bf{H}}_{{\rm{B}}{{\rm{R}},2}}}$. Let ${{\bf{H}}_k} = {\bf{h}}_k^H{{\bf{h}}_k}$ and ${{\bf{W}}_k} = {{{\bf{w}}_k}}{{\bf{w}}_k^H}$, where $\mathrm{rank} \left( {{\mathbf{W}_k}} \right) = 1,\sum\limits_{k = 1}^K {\text{Tr}\left( {{{\bf{W}}_k}} \right)}  \le {P_{\max }},{{\bf{W}}_k}\underline  \succ  0$.

In this context, Eq.~\eqref{eq:SINR}~can be rewritten as
\begin{align}
{\rm{SIN}}{{\rm{R'}}_k} = \frac{{\text{Tr}\left( {{{\bf{H}}_k}{{\bf{W}}_k}} \right)}}{{\sum\limits_{j = 1,j \ne k}^K {\text{Tr}\left( {{{\bf{H}}_k}{{\bf{W}}_j}} \right) + {\sigma ^2}} }}.
\end{align}

Hence, problem~\eqref{eq:problem 3}~as follows
\begin{equation}
\begin{aligned}
& \max_{{{\bf{W}}_k}} \quad \tau \\
& \text{s.t.} \quad
\begin{aligned}[t]
& \text{C1}'': \text{SINR}'_k \geq \tau, \\
& \text{C2}': \sum_{k=1}^K \text{Tr}(\mathbf{W}_k) \leq P_{\max}, \\
& \text{C4}: \text{rank}(\mathbf{W}_k) = 1 ,\\
& \text{C5}: \mathbf{W}_k \succeq 0.
\end{aligned}
\end{aligned}
\label{eq:problem 4}
\end{equation}

Among these constrains, $\text{C2}'$ and C5 are convex, while $\text{C1}''$ becomes convex when $\tau$ is fixed. Therefore, by relaxing the rank-1 constraint C4 and fixing the value of $\tau $, problem~\eqref{eq:problem 4}~can be transformed into a feasibility problem

\begin{equation}
\begin{aligned}
& \text{find} \quad \mathbf{W}_k\ \\
& \text{s.t.} \quad
\begin{aligned}[t]
& \text{C1}'': \text{SINR}'_k \geq \tau ,\\
& \text{C2}': \sum_{k=1}^K \text{Tr}(\mathbf{W}_k) \leq P_{\max}, \\
& \text{C5}: \mathbf{W}_k \succeq 0.
\end{aligned}
\end{aligned}
\label{eq:problem 5}
\end{equation}

The problem is now convex and can be efficiently solved using the bisection method with convex optimization tools. Specifically, the procedure begins by determining the maximum and minimum values of $\tau $ (denoted as ${\tau _{\max }}$ and ${\tau _{\min }}$), $\tau  = \left( {{\tau _{\min }} + {\tau _{\max }}} \right)/2$. If a feasible solution exists for $\tau $, set ${\tau _{\min }} = \tau $; otherwise, set ${\tau _{\max }} = \tau $. The iteration stops when ${\tau _{\max }} - {\tau _{\min }} \le \varepsilon $, where $\varepsilon $ is a tolerance value. Note that after solving for ${{{\bf{W}}_k}}$, the rank-1 constraint C4 must be verified. If this constraint fails to hold, the Gaussian approximation is applied to transform it into a residual constraint\cite{5447068}.

Based on the above, we propose a transmit beamforming optimization scheme, with the detailed procedure summarized in Algorithm 1.

\begin{algorithm}[t]
\caption{Transmit Beamforming Optimization}
\begin{algorithmic}[1]
\STATE
\textbf{Input:} The channels ${\bf{h}}_{i,k}^H$, ${{\bf{H}}_{{\rm{B}}{{\rm{R}},i}}}$, and the phase shift matrices ${{\bf{\Theta }}_1}$ and ${{\bf{\Theta }}_2}$.
\STATE
\textbf{Output:} ${{{\bf{W}}_k}}$
\STATE
\textbf{Initialize:} Set ${{\tau _{\min }}}$, ${{\tau _{\max }}}$ and $\varepsilon $, and given ${{\bf{H}}_k}$ by SDR.
\STATE
\textbf{Repeat:}
\STATE
Set $\tau  = \left( {{\tau _{\min }} + {\tau _{\max }}} \right)/2$
\STATE
Solve the feasibility program using CVX:
\begin{equation}
\begin{aligned}
& \text{find} \quad \mathbf{W}_k \\
& \text{s.t.} \quad
\begin{aligned}[t]
& \text{C1}'',\text{C2}',\text{C5}.\\
\end{aligned}
\end{aligned}
\label{eq:feasibility_problem}
\end{equation}
\STATE
If problem is feasible, set ${\tau _{\min }} = \tau $, and set ${{{\bf{W}}_k}}$ as the optimal transmit beamforming. Otherwise, set ${\tau _{\max }} = \tau $.
\STATE
\textbf{Until:} ${\tau _{\max }} - {\tau _{\min }} \le \varepsilon $
\end{algorithmic}
\end{algorithm}

\subsection{Dual-RIS-Assisted Angle Sensing}

Given the transmit beamforming vectors ${{\bf{W}}_k}$, we proceed to optimize RIS phase shift matrices ${{\bf{\Theta }}_1}$ and ${{\bf{\Theta }}_2}$, with the corresponding optimization subproblem formulated as

\begin{equation}
\begin{aligned}
& \max_{\boldsymbol{\Theta}_1, \boldsymbol{\Theta}_2} \quad \tau \\
& \mathrm{s.t.} \quad
\begin{aligned}[t]
& \mathrm{C1}'': \mathrm{SINR}'_k \geq \tau ,\\
& \mathrm{C3}: \theta_{i,n} \in \mathcal{F}.\\
\end{aligned}
\end{aligned}
\label{eq:problem 6}
\end{equation}

Due to the non-convexity introduced by discrete phase shifts, conventional convex optimization methods cannot be directly applied to solve this problem. For discrete optimization problems, global search represents the most straightforward approach. While this method is viable when the number of RIS elements is small, its computational complexity becomes prohibitive for larger-scale RIS configurations. To tackle this challenge, we consider leveraging the sensing capability of the ISAC system to facilitate the solution. Specifically, by utilizing sensed angle information, we can effectively reduce the search space of discrete phase shifts, thereby substantially lowering the computational overhead of the optimization process.

This work considers NLoS scenarios by employing dual-RIS assistance to accomplish the sensing tasks. The echo signals received by the RIS sensing elements undergo signal processing, where the two-dimensional multiple signal classification (2D-MUSIC) algorithm is employed to estimate the user's angle information\cite{9512486}.

First, we process the received echo signal of RIS $i$ as follows
\begin{equation}
\begin{aligned}
\mathbf{y}_i(t) &={{{\bf{H}}_{{{\mathop{\rm R}\nolimits}}{{\mathop{\rm U}\nolimits}}{{\mathop{\rm S}\nolimits} ,i}}}{{\bf{\Theta }}_i}{{\bf{H}}_{{\mathop{\rm B}\nolimits} {{\mathop{\rm R}\nolimits} ,i}}}{{\bf{w}}_s}{x_1}\left( t \right)}  + {{\bf{n}}_i}\left( t \right) \\
&= \sum_{k=1}^K \alpha_{3,i} \mathbf{a}_{\rm{S}}(u_{k,i},v_{k,i}) \mathbf{a}_{\rm{R}}^T(u_{k,i},v_{k,i}) \text{diag}(\boldsymbol{\theta}_i) \\
&\quad \times \alpha_{1,i} \mathbf{a}_{\rm{R}}(u_{{\rm{B}} {\rm{R}},i},v_{{\rm{B}} {\rm{R}},i}) \mathbf{b}_{\rm{B}}^T(v_{{\rm{B}} {\rm{R}},i}) \mathbf{w}_s x_1(t) + \mathbf{n}_i(t) \\
&\stackrel{\text{def}}{=} \mathbf{F}(u,v) \text{diag}(\boldsymbol{\theta}_i) \mathbf{H}(t),
\end{aligned}
\label{eq:channel_model}
\end{equation}
where
\begin{align}
\mathbf{F}(u, v) &= \Big[ \alpha_{3,i} \mathbf{a}_{\mathrm{S}}(u_{1,i}, v_{1,i}) \mathbf{a}_{\mathrm{R}}^T(u_{1,i}, v_{1,i}), \nonumber \\
&\, \cdots, \alpha_{3,i} \mathbf{a}_{\mathrm{S}}(u_{K,i}, v_{K,i}) \mathbf{a}_{\mathrm{R}}^T(u_{K,i}, v_{K,i}) \Big], \\
\mathbf{H}(t) &= \alpha_{1,i} \mathbf{a}_{\mathrm{R}}(u_{\mathrm{BR},i}, v_{\mathrm{BR},i}) \mathbf{b}_{\mathrm{B}}^T(v_{\mathrm{BR},i}) \mathbf{w}_s x_1(t).
\end{align}

Next, the received signals are stacked over $T$ snapshots as ${\bf{Y}} = \left[ {{{\bf{y}}_i}\left( 1 \right), \cdots {{\bf{y}}_i}\left( T \right)} \right]$, with their covariance matrix given by ${\bf{R}} = \frac{1}{T}{\bf{Y}}{{\bf{Y}}^H}$.

Subsequently, eigenvalue decomposition is performed on the covariance matrix:
\begin{align}
{\bf{R = ED}}{{\bf{E}}^{H}}{\bf{ = }}{{\bf{E}}_{s}}{{\bf{D}}_{s}}{\bf{E}}_{s}^{H}{\bf{ + }}{{\bf{E}}_{\text{o}}}{{\bf{D}}_{\text{o}}}{\bf{E}}_{\text{o}}^{H}.
\end{align}
where ${{\bf{E}}_{s}}\in \mathbb{C}^{M\times K}$ and ${{\bf{E}}_{o}}\in \mathbb{C}^{M\times (M-K)}$ are the eigenvectors that span the signal and noise subspaces, respectively. ${{\bf{D}}_{s}}$ is an $K\times K$ diagonal matrix whose diagonal elements contain the largest $K$ singular values. ${{\bf{D}}_{\text{o}}}$ represents an $(M-K)\times (M-K)$ diagonal matrix whose diagonal entries contain the rest eigenvalues. The eigenvalues corresponding to the signal subspace are larger than those of the noise subspace. Therefore, through eigenvalue sorting, we can decompose the covariance matrix into mutually orthogonal signal and noise subspaces.

By leveraging the orthogonality between the signal and noise subspaces, i.e., ${\bf{F}}\left( {u,v} \right)$ being orthogonal to ${{\bf{E}}_{o}}$, we perform a spectral peak search, where the largest peak corresponds to the estimated angle.
\begin{equation}
\begin{aligned}
P(u,v) &= \frac{1}{\|\mathbf{F}(u,v)\mathbf{E}_{o}\|_F^2} \\
&= \frac{1}{[\mathbf{a}(u) \otimes \mathbf{a}(v)]^H \mathbf{E}_{o} \mathbf{E}_{o}^H [\mathbf{a}(u) \otimes \mathbf{a}(v)]}.
\end{aligned}
\label{eq:power_spectrum}
\end{equation}

Based on the aforementioned formulas~\eqref{eq:steering_vector} and~\eqref{eq:steering_vector2}, the elevation and azimuth angles $\left( {{\phi _{\rm des}},{\varphi _{\rm des}}} \right)$ are computed. The phase shift for each reflective element is then determined according to the following equation\cite{9206044}
\begin{align}
\begin{array}{l}
{\theta _{y,z}^*}\\ = \bmod \left( {\left( {\left( {\sin {\phi _t}\cos {\varphi _t} + \sin {\phi _{\rm des}}\cos {\varphi _{\rm des}}} \right)\left( {y - \frac{1}{2}} \right){d_y}} \right.} \right.\\
\left. {\left. { + \left( {\sin {\phi _t}\sin {\varphi _t} + \sin {\phi _{\rm des}}\sin {\varphi _{\rm des}}} \right)\left( {z - \frac{1}{2}} \right){d_z}} \right),\left. {2\pi } \right)} \right),
\label{eq:phase compute}
\end{array}
\end{align}
where the size of each unit cell along the y axis is ${{d_y}}$ and that along the z axis is ${{d_z}}$, $\left( {{\phi _t},{\varphi _t}} \right)$ denote the elevation angle and the azimuth angle from the BS to the RIS center, and $\left( {{\phi _{\rm des}},{\varphi _{\rm des}}} \right)$ denote the elevation and azimuth angles of the sensing target direction, namely the elevation and azimuth angles from the user to the RIS.

Specifically, in the single-user scenario, Eq.~\eqref{eq:phase compute}~can be employed to compute the phase shift for each RIS element, thereby obtaining the optimal phase shift matrix ${{\bf{\Theta }}^*} = diag(\theta _{1,1}^*, \cdots ,\theta _{y,z}^*, \cdots \theta _{N_y,N_z}^*)$, where $\theta _{y,z}^*$ represents the optimal phase shift of the element located at the $y$-th row and $z$-th column of the RIS. However, in multi-user scenarios, ${{\bf{\Theta }}^*}$ cannot be directly computed with this approach. For solving the challenging problem, we propose calculating the optimal phase shifts for each user individually. As a result, the $n$-th unit of the RIS will have multiple potential optimal phase shift values. Under ideal conditions, these computed values are continuous. By quantizing their maximum value ${F_{i,n}^{\max}}$ and minimum value ${F_{i,n}^{\min}}$, the discrete phase shift range ${{\mathcal{F}}_{i,n}}$ for the $n$-th element of RIS $i$ is obtained as ${\mathcal{F}_{i,n}} = \left[ {{F_{i ,n}^{\min}},{F_{i,n}^{\max}}} \right]$. We define
\begin{align}
\begin{array}{l}
{F_{i ,n}^{\min}} = \mathop {\arg \min \left| {\theta _{\min }^* -\mathcal{F} } \right|}\limits_{         {\rm{                        }}}, \\
{F_{i,n}^{\max}} = \mathop {\arg \min \left| {\theta _{\max }^* -\mathcal{F} } \right|}\limits_{         {\rm{                        }}},
\label{eq:range compute}
\end{array}
\end{align}
where $\theta _{\min }^*$ and $\theta _{\max }^*$ are the minimum and maximum values of the optimal phase shift for the $n$-th element, respectively. The upper and lower bounds define the range of optimal phase shifts for RIS elements in multi-user scenarios. For each RIS element, the potential values of the optimal phase shift can be narrowed down from the initial range $\mathcal{F}$ to the subset ${{\mathcal{F}}_{i,n}}$.

The core steps of the proposed scheme are outlined in Algorithm 2.

\begin{algorithm}[t]
\caption{RIS-Assisted User Angle Sensing}
\begin{algorithmic}[1]
\STATE
\textbf{Initialize:} The time block is divided into $T$ snapshots.
\STATE
\textbf{Input:} The elevation angle and the azimuth angle $\left( {{\phi _t},{\varphi _t}} \right)$, the number of users $K$, the size of each unit cell $d_y$ and $d_z$.
\STATE
Stack the received echo signals from the users at the RIS sensors into a matrix, denoted as ${\bf{Y}}$.
\STATE
Estimate angles $\left( {{u_{k,i}},{v_{k,i}}} \right)$ using the 2D-MUSIC algorithm.
\STATE
Calculate angles $\left( {\phi _{\rm des}},{\varphi _{\rm des}}\right)$ using Eq.~\eqref{eq:steering_vector} and~\eqref{eq:steering_vector2}.
\STATE
Computation of all optimal phase shifts $\theta _{y,z}^*$ by~\eqref{eq:phase compute}.
\STATE
Quantify the discrete phase shift feasible space ${\mathcal{F}_{i,n}}$ by~\eqref{eq:range compute}.
\STATE
\textbf{Output:} ${\mathcal{F}_{i,n}}$.
\end{algorithmic}
\end{algorithm}

\subsection{Sensing-Aided Discrete Phase Shift Optimization}

For the multi-user ISAC system, the formulated optimization problem~\eqref{eq:aligned_problem}~aims to maximize the minimum user SINR, the optimal solution for multiple users must necessarily fall within this value range. By strategically narrowing the constraint boundaries through sensing-assisted refinement, we can significantly enhance computational efficiency.

For the optimization subproblem~\eqref{eq:problem 6}, we investigate RIS phase shift design strategies under tightened discrete phase constraints across varying RIS size.

\subsubsection{Small-scale RIS}

Given the beamforming vector $\mathbf{W}_k$ and RIS $2$ phase shift matrix $\mathbf{\Theta}_2$, the optimization subproblem~\eqref{eq:problem 6}~can be reduced to
\begin{equation}
\begin{aligned}
& \underset{\mathbf{\Theta}_1}{\mathrm{max}}  \quad \tau \\
& \mathrm{s.t.} \quad
\begin{aligned}[t]
& \mathrm{C1}'': \mathrm{SINR}'_k \geq \tau,  \\
& \mathrm{C6}: \theta_{1,n} \in \mathcal{F}_{1,n}.
\end{aligned}
\end{aligned}
\label{eq:ris_optimization}
\end{equation}

In this case, the number of RIS elements $N$ is relatively small. Furthermore, as the range of discrete phase shifts $\mathcal{F}$ is substantially reduced, it becomes computationally feasible to solve the problem through direct global search. Based on the constraint set $\mathcal{F}_{1,n}$ for RIS $1$ elements, all possible phase shift matrix configurations $\boldsymbol{\chi}_1\triangleq\{ {{{\bf{\Theta }}_{1,1}},{{\bf{\Theta }}_{1,2}}, \cdots ,{{\bf{\Theta }}_{1,n}}, \cdots }\}$ can be computed. Then, by fixing the value of $\tau$ and employing the bisection method, the problem can be transformed into a feasibility problem
\begin{equation}
\begin{aligned}
& \text{find} \quad \mathbf{\Theta}_1 \\
& \text{s.t.} \quad
\begin{aligned}[t]
& \text{C7}: f_s(\mathbf{\Theta}_1) \geq \tau ,\\
& \text{C8}: \mathbf{\Theta}_1 \in \boldsymbol{\chi}_1,
\end{aligned}
\end{aligned}
\label{eq:problem 7}
\end{equation}
where ${f_s}({\bf{\Theta}}_1) = \min {\rm{SINR}}{{\rm{'}}_k}$. Through global search, we can obtain the optimal discrete solution ${\bf{\Theta }}_1^*$. On the basis of the foregoing derivation and analysis, we summarize the global search-based sensing-assisted phase shift matrix optimization scheme in Algorithm 3. Similarly, given $\mathbf{W}_k$ and $\mathbf{\Theta}_1$, the optimal discrete phase shift matrix $\mathbf{\Theta}_2$ for RIS $2$ can be derived.

Moreover, by leveraging sensing-refined constraint ranges and performing an global search, the optimal solution of subproblem~\eqref{eq:problem 7}~can be guaranteed, making this approach particularly advantageous for scenarios demanding low latency and high precision.

\begin{algorithm}[t]
\caption{Sensing-Based Global Search for Phase Shift Matrix Optimization}
\begin{algorithmic}[1]
\STATE
\textbf{Input:} The channels ${\bf{h}}_{i,k}^H$, ${{\bf{H}}_{{\rm{B}}{{\rm{R}},i}}}$, the given phase shift matrix ${{\bf{\Theta }}_2}$, the transmit beamforming ${{{\bf{W}}_k}}$, the received signal ${{\bf{y}}_i}\left( t \right)$ and the elevation angle and the azimuth angle from the BS to the RIS center $\left( {{\phi _t},{\varphi _t}} \right)$.
\STATE
\textbf{Output:} ${{\bf{\Theta }}_1}$
\STATE
\textbf{Initialize:} Set a tolerance value $\varepsilon $, ${\tau _{\min }}=0$ and ${\tau _{\max }}=10$, give ${{\bf{H}}_k}$ by SDR.
\STATE
The discrete phase-shift constraints ${\mathcal{F}_{1,n}}$ are obtained using Algorithm 2.
\STATE
\textbf{Repeat:}
\STATE
Set $\tau  = \left( {{\tau _{\min }} + {\tau _{\max }}} \right)/2$
\STATE
Solve the feasibility program using global search method:
\begin{equation}
\begin{aligned}
& \text{find} \quad \mathbf{\Theta}_1 \\
& \text{s.t.} \quad
\begin{aligned}[t]
& \text{C7}, \text{C8}. \\
\end{aligned}
\end{aligned}
\label{eq:feasibility_problem}
\end{equation}
\STATE
If problem is feasible, set ${\tau _{\min }} = \tau $, and set $\mathbf{\Theta}_1$ as the optimal phase shift matrix. Otherwise, set ${\tau _{\max }} = \tau $.
\STATE
Update ${{\bf{H}}_k}$
\STATE
\textbf{Until:} ${\tau _{\max }} - {\tau _{\min }} \le \varepsilon. $
\end{algorithmic}
\end{algorithm}

\subsubsection{Large-scale RIS}

 The above Algorithm 3 typically achieves excellent performance in small-scale RIS systems. However, as the number of RIS  elements $N$ increases, it incurs prohibitive computational complexity. To address this, a lower-complexity algorithm is essential for optimizing discrete phase shifts in large-scale RIS systems. Leveraging sensing-assisted constraint tightening, this subsection employs a one-dimensional search method to overcome the scalability challenge. Since the phase shift designs for both RIS 1 and RIS 2 follow similar principles, this subsection focuses on optimizing the discrete phase shifts $\mathbf{\Theta}_1$ for RIS 1. The same methodology can be symmetrically applied to derive $\mathbf{\Theta}_2$ for RIS 2. Given the beamforming vector $\mathbf{W}_k$ and phase shift matrix $\mathbf{\Theta}_2$, the original optimization problem~\eqref{eq:aligned_problem}~can be simplified as
\begin{equation}
\begin{aligned}
& \max_{\mathbf{\Theta}_1}  \min_{k \in \mathcal{K}} \quad \text{SINR}_k \\
& \text{s.t.} \quad
\begin{aligned}[t]
& \text{C1}: \text{SINR}_k \geq \gamma,\\
& \text{C6}: \theta_{1,n} \in \mathcal{F}_{1,n}, \forall n \in \left\{ {1, \cdots ,N} \right\}.
\end{aligned}
\end{aligned}
\label{eq:problem 8}
\end{equation}
Let $\tau = \mathop {{\rm{ min}}}\limits_{k \in \mathcal{K}} {\rm{ SIN}}{{\rm{R}}_k}$, the max-min fairness problem~\eqref{eq:problem 8}~can be reformulated as the following maximization problem
\begin{equation}
\begin{aligned}
& \max_{\theta_{1,n}} \quad \tau \\
& \text{s.t.} \quad
\begin{aligned}[t]
& \text{C9}: \tau \geq \gamma ,\\
& \text{C6}: \theta_{1,n} \in \mathcal{F}_{1,n}.
\end{aligned}
\end{aligned}
\label{eq:phase_optimization}
\end{equation}

This approach transforms the optimization of phase shift matrix $\mathbf{\Theta}_1$ into the optimization of element-wise phase shift $\theta_{1,n}$. Specifically, by fixing $N-1$ phase shifts in RIS 1 except for element $n$, a one-dimensional search over $\theta_{1,n}$ on the constrained set $ \mathcal{F}_{1,n}$ is performed in each iteration to identify the optimal $\theta_{1,n}^*$, i.e., $\theta _{1,n}^* = \arg \mathop {\max }\limits_{{\theta _{1,n}} \in {\mathcal{F}_{1,n}}} \tau$.

Based on this, we employ an alternating iteration method to obtain suboptimal solutions for the another phase shifts. Each time a suboptimal solution is obtained, the corresponding phase shift value in the phase shift matrix must be updated. Until $\tau$ no longer increases, the suboptimal phase shifts of each element constitute the suboptimal phase shift matrix of RIS $1$. The core steps of the proposed scheme are outlined in Algorithm 4. Similarly, we can obtain the optimal discrete phase shift matrix ${{\bf{\Theta }}_2}$ given the transmit beamforming ${{{\bf{W}}_k}}$ and the phase shift matrix ${{\bf{\Theta }}_1}$.

Owing to the narrowed range of discrete phase shift constraints, this approach achieves complexity reduction compared to conventional one-dimensional search methods. This lower complexity makes the method applicable to both high-resolution phase configurations and large-scale RIS systems.

\begin{algorithm}[t]
\caption{Sensing-Based One-Dimensional Search for Phase Shift Matrix Optimization}
\begin{algorithmic}[1]
\STATE
\textbf{Input:} The channels ${\bf{h}}_{i,k}^H$, ${{\bf{H}}_{{\rm{B}}{{\rm{R}},i}}}$, the given phase shift matrix ${{\bf{\Theta }}_2}$, the transmit beamforming ${{{\bf{W}}_k}}$, the received signal ${{\bf{y}}_i}\left( t \right)$ and the elevation angle and the azimuth angle from the BS to the RIS center $\left( {{\phi _t},{\varphi _t}} \right)$.
\STATE
\textbf{Output:} ${{\bf{\Theta }}_1}$
\STATE
\textbf{Initialize:} Set a tolerance value $\varepsilon $, the initial iteration number $l$ = 0, the maximum number of iterations $L$ = 1000, ${\tau _{\min }}=0$ and ${\tau _{\max }}=10$.
\STATE
The discrete phase-shift constraints ${\mathcal{F}_{1,n}}$ are obtained using Algorithm 2.
\STATE
\textbf{Repeat:}
\STATE
Set $\tau  = \left( {{\tau _{\min }} + {\tau _{\max }}} \right)/2$, $l$ = $l$ + 1.
\STATE
Solve the feasibility program using one-dimensional search method:
\begin{equation}
\begin{aligned}
& \max_{\theta_{1,n}} \quad \tau \\
& \text{s.t.} \quad
\begin{aligned}[t]
& \text{C9}, \text{C6}.  \\
\end{aligned}
\end{aligned}
\label{eq:maxmin_problem}
\end{equation}
With all phase shifts, except for ${\theta_{1,n}}$, are fixed, set $\theta_{1,n}^* = \arg \mathop {\max }\limits_{{\theta_{1,n}} \in {\mathcal{F}_{1,n}}} \tau $ as the suboptimal phase shift.
\STATE
Update ${\theta_{1,n}} = \theta_{1,n}^*$ and choose a different phase shift as the optimization variable in the next iteration.
\STATE
\textbf{Until:} $l$ = $L$ or $\mathop {\max }\limits_{{\theta_{1,n}} \in {\mathcal{F}_{1,n}}} \tau^{l} - \mathop {\max }\limits_{{\theta_{1,n}} \in {\mathcal{F}_{1,n}}} \tau^{l-1}<\varepsilon .$
\end{algorithmic}
\end{algorithm}

\subsection{Overall Algorithm and Complexity Analysis}
Based on Algorithms 1, 2, 3 and 4, the scheme for solving the original problem~\eqref{eq:aligned_problem}~is summarized as Algorithm 5, where $N_{th}$ denotes the threshold for determining the RIS size. Specifically, we first introduce an auxiliary variable $\tau$ to represent the minimum SINR, and then decompose the original problem~\eqref{eq:aligned_problem}~into three subproblems , which are subsequently addressed using an alternating optimization approach. Using Algorithm 1, by giving the phase shift matrices $\mathbf{\Theta}_1$ and $\mathbf{\Theta}_2$, the BS transmit beamforming vectors ${{{\bf{W}}_k}}$ can be optimized. Before solving the discrete phase shift optimization subproblem~\eqref{eq:problem 6}, the user's angle information is first estimated by utilizing the echo signals received at the RISs. Subsequently, the relatively optimal phase shift ${\theta _{y,z}^*}$ is calculated to reduce the feasible range of the discrete phase shifts $\mathcal{F}$. Based on these discrete constraints $\mathcal{F}_{i,n}$, either Algorithm 3 or Algorithm 4 is selected to solve the discrete phase shift matrix $\mathbf{\Theta}_i$ of the RIS $i$, depending on the RIS size.

The computational complexity of Algorithm 1 is primarily determined by the complexity of solving the SDR problem, which is typically addressed using interior-point methods. The computational complexity of an SDR problem with $p$ constraints involving $q \times q$ positive semi-definite matrices can be expressed as $\mathcal{O}\left( {p{q^{3.5}} + {p^2}{q^{2.5}} + {p^3}{q^{0.5}}} \right)$. In Algorithm 1, the number of SDR constraints $p$ is ${2K + 1}$, and the dimension of the positive semi-definite matrices $q$ is ${N_t}$. Therefore, the computational complexity of Algorithm 1 can be expressed as $\mathcal{O}\left( {(2K + 1)N_t^{3.5} + {{(2K + 1)}^2}N_t^{2.5} + {{(2K + 1)}^3}N_t^{0.5}} \right)$.

The computational complexity of Algorithm 3 is primarily determined by the complexity of the global search method. Under the original phase shift constraints, if an global search is applied, the corresponding computational complexity would be $\mathcal{O}((2^b)^N)$. In Algorithm 3, by utilizing the reduced range of discrete phase shift constraints, the computational complexity becomes the product of $N$ constraint sizes $\mathcal{O}\left(\prod_{n=1}^N |{F}_n|\right)$. Here, ${F_n}$ represents the number of discrete values in the reduced phase shift constraint set ${{\mathcal{F}}_{1,n}}$, which is significantly smaller than ${2^b}$.

The computational complexity of Algorithm 4 is likewise dominated by the one-dimensional search method. Under the original phase shift constraints, the computational complexity using the one-dimensional search method is ${2^b}$. In Algorithm 4, the computational complexity of the search-based method reduces to $\mathcal{O}\left( {{F_1} + {F_2} +  \cdots  + {F_N}} \right)$, where each ${F_n}$ is significantly smaller than ${2^b}$, resulting in a substantially lower computational complexity compared to the original one-dimensional search approach.

Therefore, the complexity of Algorithm 5 is $\mathcal{O}( I_5( {(2K + 1)N_t^{3.5} + {{(2K + 1)}^2}N_t^{2.5} + {{(2K + 1)}^3}N_t^{0.5}} +\prod_{n=1}^N |{F}_n|)) $ in small-scale RIS scenarios and $\mathcal{O}( I_5( {(2K + 1)N_t^{3.5} + {{(2K + 1)}^2}N_t^{2.5} + {{(2K + 1)}^3}N_t^{0.5}} +\sum\nolimits_{n = 1}^N {\left| {{F_n}} \right|} )) $ in large-scale RIS scenarios.

\begin{algorithm}[t]
\caption{Sensing-Based Joint Beamforming}
\begin{algorithmic}[1]
\STATE
\textbf{Input:} The channels ${\bf{h}}_{i,k}^H$, ${{\bf{H}}_{{\rm{B}}{{\rm{R}},i}}}$, initial iteration number $l$ = 1, set the maximum number of iterations $L$ = 100 and set a tolerance value $\varepsilon $. Set the values of ${\tau _{\min }}=0$, ${\tau _{\max }}=10$ and the threshold $N_{th}$.
\STATE
\textbf{Repeat:}
\STATE
Set $\tau  = \left( {{\tau _{\min }} + {\tau _{\max }}} \right)/2$
\STATE
For the given phase shift matrices ${{\bf{\Theta }}_1}$ and ${{\bf{\Theta }}_2}$, solve subproblem~\eqref{eq:problem 5}~using Algorithm 1, and then obtain the optimal transmit beamforming ${{{\bf{W}}_k}}$.
\STATE
\textbf{IF} $N<N_{th}$
\STATE
The discrete phase shift optimization subproblem~\eqref{eq:problem 6} is solved by Algorithm 3.
\STATE
\textbf{ELSE} the subproblem~\eqref{eq:problem 6} is solved by Algorithm 4.
\STATE
Update ${{\bf{\Theta }}_1}$ and ${{\bf{\Theta }}_2}$, $l=l+1$.
\STATE
\textbf{Until:} $l$ = $L$ or ${\tau}^{l}-\tau^{l-1}<\varepsilon .$
\STATE
\textbf{Output:} ${{{\bf{W}}^{l}_k}}$, ${{\bf{\Theta }}^{l}_1}$ and ${{\bf{\Theta }}^{l}_2}$
\end{algorithmic}
\end{algorithm}

\section{SIMULATION RESULTS}
For the dual-RIS-assisted multi-user ISAC system, numerical simulation results validate the effectiveness of the proposed joint beamforming algorithm. Meanwhile, for ease of description, the two sensing-based RIS discrete phase shift design algorithms, namely Algorithm 3 and Algorithm 4, are hereafter referred to as the GS-Algorithm and 1D-Algorithm, respectively.

\subsection{Simulation Parameters}

In the simulation setup, a 3D coordinate system is considered, with the BS located at (10 m, 0 m, 10 m), and two RISs positioned on the plane $x=0$, respectively 45 m and 50 m from the BS, while users on the same side of the two RISs are randomly distributed 40-70 m away from them. Meanwhile, we adopt the distance-dependent path loss model $\alpha  = {C_0}{\left( {\frac{q}{{{q_r}}}} \right)^{ - \kappa }}$\cite{9246254}, where the path loss exponents from the BS to the RIS, the RIS to the users and RIS receiving echo signals from users are set as: $\kappa_1=2.2, \kappa_2=2.5$ and $\kappa_3=2.5$. The path loss at $q_r=$ 1 m is set to $C_0$ = 30 dB. The BS is equipped with $N_t=6$ transmit antennas, with a transmit power of $P_{\rm max}$ = 30 dBm. Each RIS comprises $N=16$ reflective elements and $M=64$ sensing elements, with the adjacent element spacing set to $d=d_y=d_z=$ 0.01 m. Moreover, we set ${\tau _{\min }}=0$ and ${\tau _{\max }}=10$. The number of users is $K=4$, the number of snapshots is $L=1000$, and the noise power is -40 dBm. To better demonstrate the effectiveness of the proposed sensing-based beamforming algorithm, the following two benchmark schemes are provided for comparison.

\begin{itemize}
  \item \textbf {AO-Continuous Algorithm}: The deployment of the BS and RISs remains unchanged, and the RIS phase shifts are assumed to be continuous. The beamforming optimization method for the BS is consistent with the proposed scheme. Since the continuous phase shift constraint of the RIS is already convex, the problem is transformed using SDR and solved with the same method as described earlier. An alternating optimization approach is then applied to the beamforming of the BS and RIS.
  \item \textbf {Discrete Quantization Algorithm}: Building upon the AO-Continuous Algorithm, the quantization of the optimal phase shift for each RIS element is performed by selecting the discrete value that most closely approximate its optimal continuous counterparts.
\end{itemize}

On this basis, we evaluate the max-min SINR performance of the proposed beamforming schemes based on the GS-Algorithm and 1D-Algorithm, respectively. It should be noted that the SINR values are directly expressed as power ratios without logarithmic scaling.

\subsection{Analysis of GS Algorithm}
\begin{figure}[t]
    \centering
    \includegraphics[ width=1\linewidth]{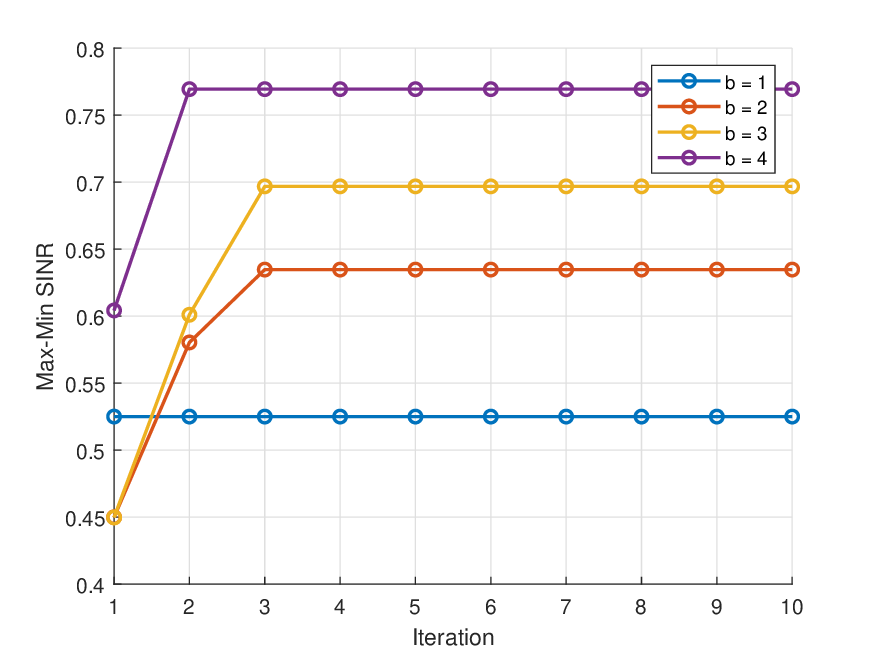}
    \caption{The convergence of the GS algorithm under different $b$ values.}
\end{figure}

This subsection evaluates the performance of the proposed joint beamforming scheme based on GS Algorithm. We begin by evaluating the convergence performance of the proposed GS algorithm based on the max-min SINR criterion, as shown in Fig. 3, which displays the convergence curves for different phase-shift bit numbers $b$. The vertical axis represents the minimum SINR of the users. Intuitively, the algorithm converges within three iterations for all values of $b$. When $b=1$, convergence can be achieved with only one outer loop iteration. This is because only two discrete phase-shift values are available initially, and after refining the feasible set via sensing-assisted calculation, the narrowed set usually only contains one feasible value, resulting in immediate convergence. Moreover, the results indicate that system performance improves as $b$ increases. This trend occurs because a larger $b$ exponentially expands the number of possible discrete phase-shift values, thereby approaching the optimal continuous phase shift. In the limit as $b\rightarrow\infty$, the discrete phase-shift model approaches the ideal continuous phase-shift scenario. Additionally, the algorithm achieves a relatively high max-min SINR after the first iteration. For $b=1$ and $b=2$, the max-min SINR reaches approximately 0.45 after the initial iteration, while for $b=4$, it exceeds 0.6. This demonstrates that by narrowing the discrete phase-shift search space, suboptimal values are effectively eliminated, thereby accelerating the convergence speed of the proposed algorithm.

\begin{figure}[t]
    \centering
    \includegraphics[width=1\linewidth]{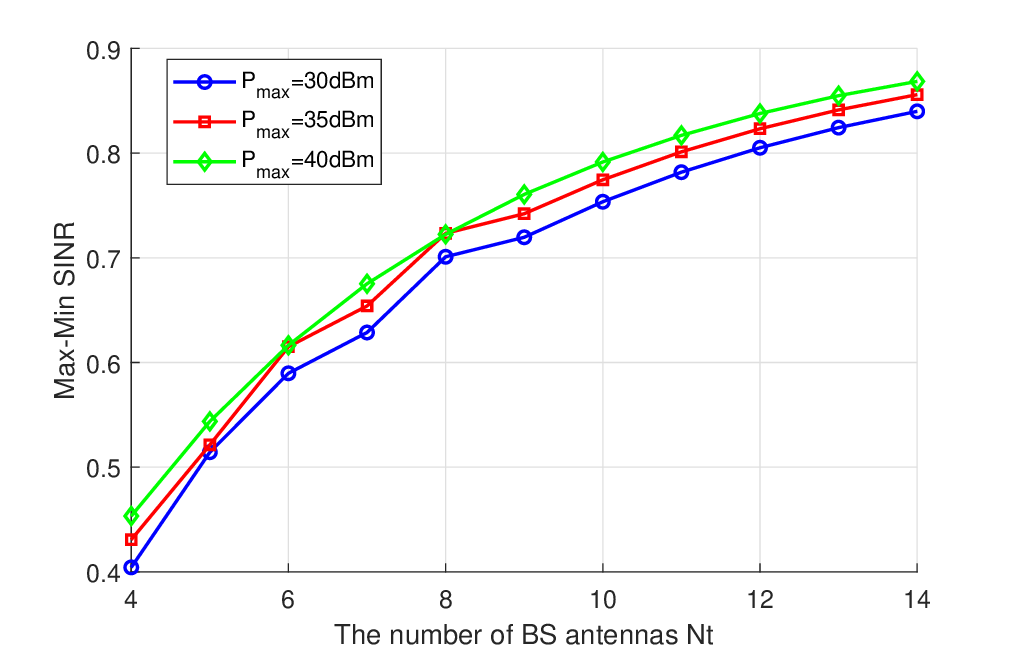}
    \caption{Fairness varies with the number of BS antennas $N_t$ and the transmission power $P_{\rm max}$.}
\end{figure}

We then evaluate the proposed scheme performance under the GS algorithm with different BS transmit power levels versus the number of BS transmit antennas $N_t$, as shown in Fig. 4. It can be intuitively observed that the communication performance in terms of max-min SINR improves as $N_t$ increases. Specifically, when the number of antennas is small (e.g., $N_t<8)$, each additional antenna brings a significant performance gain, increasing max-min SINR by more than 0.1 per antenna. Beyond this point, the growth rate gradually slows down and the curve tends to saturate. This trend occurs because a larger $N_t$ enhances spatial multiplexing gain, though this improvement is subject to diminishing returns. From a vertical perspective, for a fixed $N_t$, a higher maximum transmit power $P_{\rm max}$ leads to better performance. Notably, when $N_t=6$ and $N_t=8$, the curves for $P_{\rm max}=$ 30 dBm and $P_{\rm max}=$ 35 dBm exhibit noticeable fluctuations. This is attributed to the use of SDR in the proposed algorithm to tackle the non-convex problem. The Gaussian randomization process may produce a solution with a rank greater than one, leading to max-min SINR instability. Overall, however, for a given $N_t$, the improvement in max-min SINR remains roughly proportional to the increase in $P_{\rm max}$.

\begin{figure}[t]
    \centering
    \includegraphics[ width=1\linewidth]{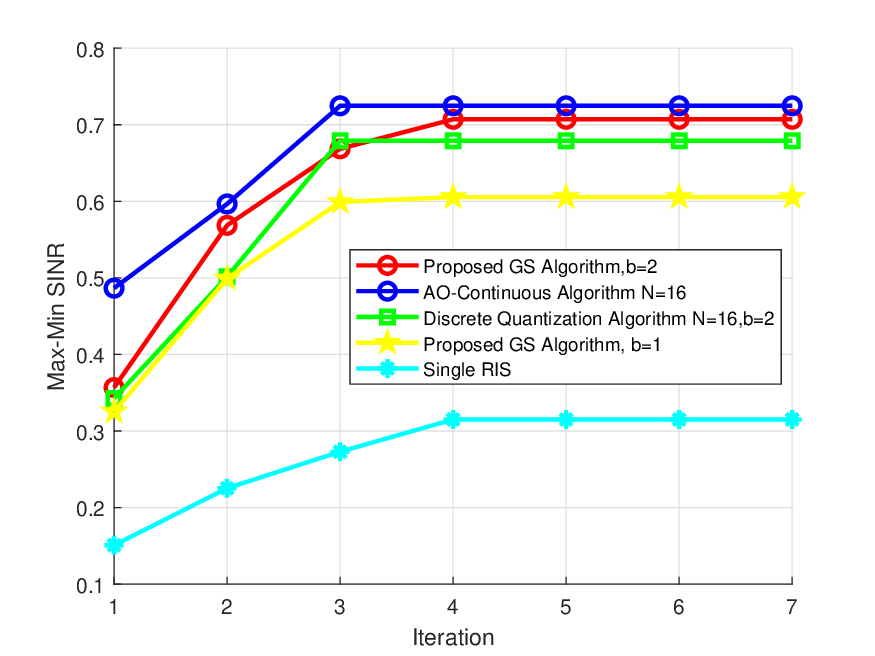}
    \caption{Comparison of the max-min SINR of different algorithms.}
\end{figure}

Next, we compare the max-min SINR performance of the proposed GS algorithm-based joint beamforming scheme with the two benchmarks. Fig. 5 shows the convergence of the three algorithms under different configurations, with all other parameters held constant except as indicated. As observed, the continuous phase shift algorithm achieves the highest max-min SINR after convergence. Notably, for $b=2$, the proposed algorithm attains the second-highest max-min SINR, approaching the performance of the ideal continuous case and outperforming the discrete quantization algorithm under the same phase shift resolution. When $b=1$, the proposed algorithm converges to a value of 0.6, showing only a marginal gap compared to the discrete quantization result at $b=2$. In terms of convergence speed, the proposed algorithm requires a similar number of iterations as the continuous phase shift benchmark. We also evaluate performance under different RIS deployment configurations. A single-RIS scenario is considered, where the total number of elements matches the sum of elements in the dual-RIS setup, with all other settings unchanged. The single-RIS configuration achieves a converged max-min SINR of only 0.3, which is more than two times lower than the 0.7 achieved by the dual-RIS system. This result clearly demonstrates the advantage of distributed multi-RIS deployments over a single RIS with an equivalent number of elements.

\begin{figure}[t]
    \centering
    \includegraphics[ width=1\linewidth]{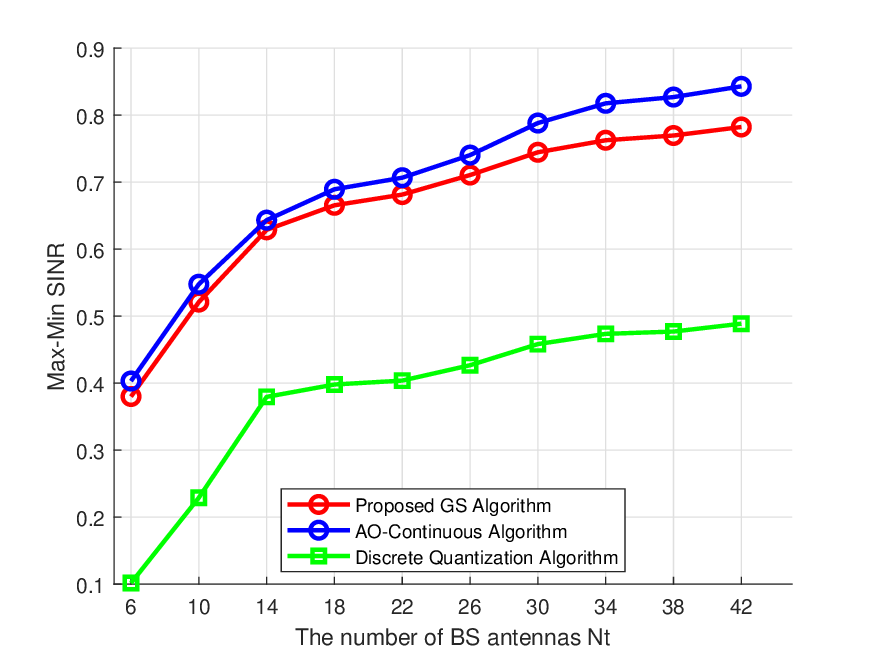}
    \caption{Comparison of different algorithms versus BS antenna numbers $N_t$.}
\end{figure}

We also compare the performance of the three algorithms as the number of BS antennas $N_t$ increases, as illustrated in Fig. 6, where $b=1$ and $N=16$. It can be observed intuitively that the user max-min SINR improves with $N_t$, due to the increased degrees of freedom offered by more antennas. The rate of improvement, however, slows after $N_t=14$. Furthermore, the proposed GS algorithm achieves performance close to that of the ideal continuous phase shift benchmark. When the number of antennas is small, the performance gap is approximately 0.02; this gap gradually increases to about 0.06 as $N_t$ grows. In comparison with the discrete quantization algorithm-which also accounts for discrete phase shifts-GS algorithm demonstrates a clear and significant advantage. At $N_t=6$, the performance gap is around 0.3. As $N_t$ increases, the gap widens to approximately 0.4, representing a near twofold difference. These results confirm the substantial superiority of the proposed GS algorithm in practical scenarios involving discrete phase shifts.

\subsection{Analysis of 1D Algorithm}

\begin{figure}[t]
    \centering
    \includegraphics[width=1\linewidth]{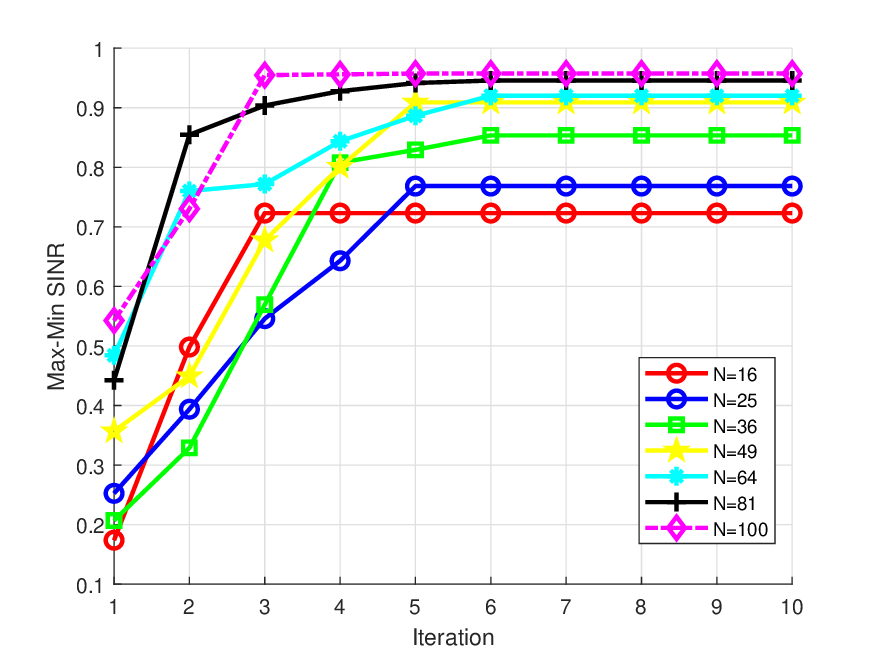}
    \caption{Convergence of the 1D algorithm under different $N$ values}
\end{figure}

This subsection evaluates the performance of the proposed joint beamforming scheme based on the 1D Algorithm. Fig. 7 illustrates the convergence performance of 1D algorithm, with each curve representing a different number of RIS elements $N$. As shown, some curves converge within 3 iterations, while all achieve convergence by the 5th iteration, demonstrating the effectiveness and robustness of the algorithm. Although the initial max-min SINR values after the first iteration vary significantly-ranging from below 0.1 to above 0.6 across configurations-the values increase rapidly in the second and third iterations, approaching convergence. This behavior occurs because the one-dimensional search optimizes RIS phase shifts sequentially. Even when initial values are suboptimal, the algorithm requires only one or two iterations to escape poor local solutions, indicating that the narrowed discrete phase-shift search space significantly enhances the convergence of 1D algorithm. Furthermore, the converged max-min SINR improves as the number of RIS elements $N$ increases. This is attributed to the greater flexibility in reflection and higher spatial degrees of freedom afforded by a larger RIS. However, this growth trend slows when $N$ reaches a certain threshold (e.g., $N=64$), due to the inherent upper limit on the received energy at the RIS elements under a fixed BS transmit power budget.

\begin{figure}[t]
    \centering
    \includegraphics[ width=1\linewidth]{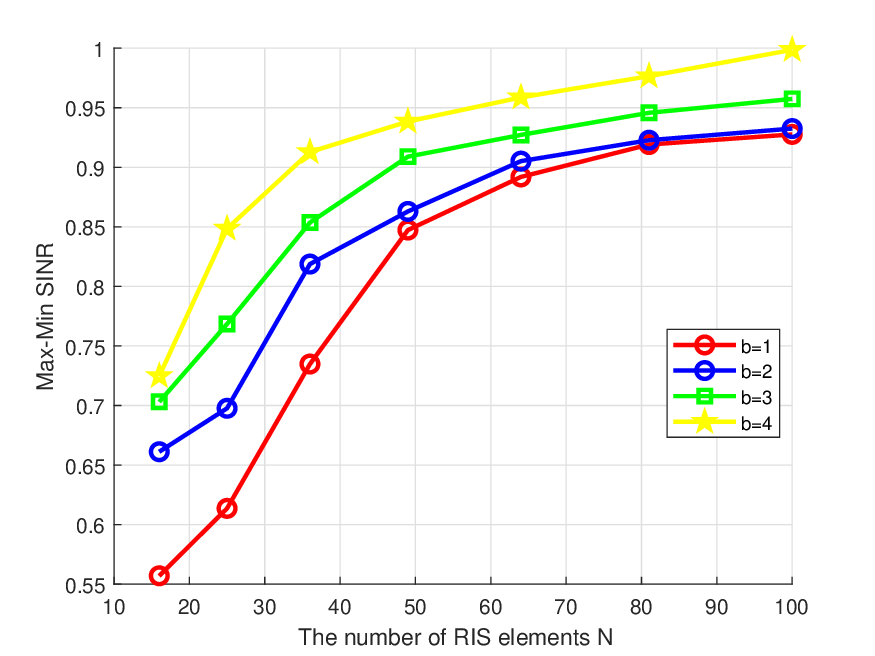}
    \caption{Comparison of the max-min SINR versus the number of RIS elements $N$ and bit resolutions $b$.}
\end{figure}

Fig. 8 shows the performance of the proposed 1D algorithm based scheme as a function of the number of RIS elements $N$ under different values of $b$. It is evident that the max-min SINR increases with $N$, owing to the enhanced spatial multiplexing gain and the increased energy captured by the RIS elements. The most significant improvement occurs when $N<49$, beyond which the growth rate gradually slows due to the upper limit on the received energy at the RIS under a fixed BS transmit power. From a vertical perspective, the performance gap between different $b$ values is more pronounced when $N$ is small. For instance, at $N=25$, the gap between adjacent $b$ values ranges from 0.05 to 0.1, while the difference between $b=1$ and $b=4$ reaches 0.25. However, when $N$ exceeds 49, this gap narrows substantially. At $N=64$, the performance difference between $b=4$ and $b=1$ is only about 0.05. These results indicate that in large-scale RIS configurations, the performance gain from increasing $N$ diminishes the impact of phase-shift resolution $b$. In contrast, both $N$ and $b$ play critical roles in small-scale RIS systems.

\begin{figure}[t]
    \centering
    \includegraphics[ width=1\linewidth]{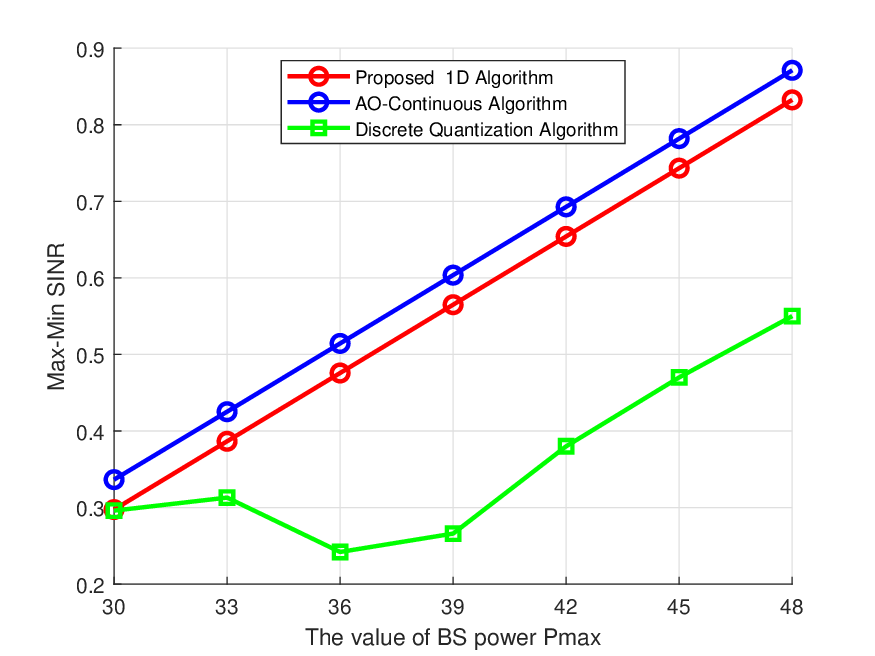}
    \caption{Performance comparison of the three different algorithms.}
\end{figure}

Similar to the GS algorithm based scheme, we compare the max-min SINR performance of the 1D algorithm based scheme with the two benchmarks as the BS transmit power $P_{\rm max}$ increases, as shown in Fig. 9. The proposed 1D algorithm exhibits a nearly linear growth curve with $P_{\rm max}$, similar to the continuous algorithm, indicating a proportional relationship. The performance gap between 1D algorithm and the continuous benchmark remains consistently small (approximately 0.03) and stable across different power levels. In contrast, while 1D Algorithm and the discrete quantization algorithm show similar initial performance at low power levels, their gap widens significantly as $P_{\rm max}$ increases. At $P_{\rm max}=$ 42 dBm, the performance difference reaches 0.3-a substantial margin. Furthermore, the discrete quantization algorithm exhibits a performance degradation when $P_{\rm max}$ increases from 33 dBm to 36 dBm, and its performance at 39 dBm remains lower than that at 30 dBm. This counterintuitive behavior stems from the fact that quantizing continuous phase shifts into discrete values for each RIS element individually leads to error accumulation. As a result, higher transmit power exacerbates the performance loss. According to the number of phase-shift level formula $2^{b}$, this effect becomes more severe when $b$ is smaller.

\section{CONCLUSION}

In this paper, to better leverage the inherent coexistence of sensing and communication in ISAC systems, we constructed a dual-RIS-assisted ISAC framework. By integrating sensing elements on the RIS, we utilized sensing-acquired information to assist the design of discrete phase shifts, thereby enhancing communication performance. To address the challenge of designing discrete phase shifts in multi-user scenarios, we introduced a phase shift calculation method from single-user scenarios to formulate the phase shift matrix, effectively narrowing the feasible set of discrete phase shift constraints. For the difficulty in optimizing discrete phase shifts, we proposed two low-complexity algorithms based on the refined phase shift design: GS algorithm for small-scale RIS configurations, which significantly reduces search complexity by leveraging the narrowed solution space; and a sensing-based 1D algorithm for large-scale RIS scenarios, which lowers the complexity from exponential to logarithmic order compared to conventional search methods. Simulation results demonstrated that both proposed algorithms outperform traditional discrete quantization approaches and achieved performance close to that of ideal continuous phase shifts. The global search algorithm satisfies high-performance demands, while the one-dimensional method, with its complexity invariant to the number of RIS elements $N$, is more suitable for low-complexity requirements. Additionally, we verify that multi-RIS deployments yield greater performance gains compared to a single-RIS setup. In summary, this work provides a practical and efficient solution for RIS-assisted ISAC systems in real-world applications.

\bibliographystyle{IEEEtran} 
\bibliography{references}

\begin{thebibliography}{10}
\providecommand{\url}[1]{#1}
\csname url@samestyle\endcsname
\providecommand{\newblock}{\relax}
\providecommand{\bibinfo}[2]{#2}
\providecommand{\BIBentrySTDinterwordspacing}{\spaceskip=0pt\relax}
\providecommand{\BIBentryALTinterwordstretchfactor}{4}
\providecommand{\BIBentryALTinterwordspacing}{\spaceskip=\fontdimen2\font plus
\BIBentryALTinterwordstretchfactor\fontdimen3\font minus
  \fontdimen4\font\relax}
\providecommand{\BIBforeignlanguage}[2]{{%
\expandafter\ifx\csname l@#1\endcsname\relax
\typeout{** WARNING: IEEEtran.bst: No hyphenation pattern has been}%
\typeout{** loaded for the language `#1'. Using the pattern for}%
\typeout{** the default language instead.}%
\else
\language=\csname l@#1\endcsname
\fi
#2}}
\providecommand{\BIBdecl}{\relax}
\BIBdecl

\bibitem{9737357}
F.~Liu, Y.~Cui, C.~Masouros, J.~Xu, T.~X. Han, Y.~C. Eldar, and S.~Buzzi,
  ``Integrated sensing and communications: Toward dual-functional wireless
  networks for 6{G} and beyond,'' \emph{IEEE Journal on Selected Areas in
  Communications}, vol.~40, no.~6, pp. 1728--1767, 2022.

\bibitem{9585321}
J.~A. Zhang, M.~L. Rahman, K.~Wu, X.~Huang, Y.~J. Guo, S.~Chen, and J.~Yuan,
  ``Enabling joint communication and radar sensing in mobile networks-{A}
  survey,'' \emph{IEEE Communications Surveys \& Tutorials}, vol.~24, no.~1,
  pp. 306--345, 2022.

\bibitem{7746569}
A.~Hassanien, M.~G. Amin, Y.~D. Zhang, and F.~Ahmad, ``Signaling strategies for
  dual-function radar communications: An overview,'' \emph{IEEE Aerospace and
  Electronic Systems Magazine}, vol.~31, no.~10, pp. 36--45, 2016.

\bibitem{9705498}
A.~Liu, Z.~Huang, M.~Li, Y.~Wan, W.~Li, T.~X. Han, C.~Liu, R.~Du, D.~K.~P. Tan,
  J.~Lu, Y.~Shen, F.~Colone, and K.~Chetty, ``A survey on fundamental limits of
  integrated sensing and communication,'' \emph{IEEE Communications Surveys \&
  Tutorials}, vol.~24, no.~2, pp. 994--1034, 2022.

\bibitem{8999605}
F.~Liu, C.~Masouros, A.~P. Petropulu, H.~Griffiths, and L.~Hanzo, ``Joint radar
  and communication design: Applications, state-of-the-art, and the road
  ahead,'' \emph{IEEE Transactions on Communications}, vol.~68, no.~6, pp.
  3834--3862, 2020.

\bibitem{10422712}
Q.~Xue, C.~Ji, S.~Ma, J.~Guo, Y.~Xu, Q.~Chen, and W.~Zhang, ``A survey of beam
  management for mmwave and {THz} communications towards 6{G},'' \emph{IEEE
  Communications Surveys \& Tutorials}, vol.~26, no.~3, pp. 1520--1559, 2024.

\bibitem{11060909}
H.~Zhang, X.~Huang, X.~Guo, S.~He, C.~Gu, Y.~Shu, and J.~Chen, ``Terahertz
  sensing, communication, and networking: A survey,'' \emph{IEEE Transactions
  on Network Science and Engineering}, pp. 1--23, 2025.

\bibitem{9140329}
M.~Di~Renzo, A.~Zappone, M.~Debbah, M.-S. Alouini, C.~Yuen, J.~de~Rosny, and
  S.~Tretyakov, ``Smart radio environments empowered by reconfigurable
  intelligent surfaces: How it works, state of research, and the road ahead,''
  \emph{IEEE Journal on Selected Areas in Communications}, vol.~38, no.~11, pp.
  2450--2525, 2020.

\bibitem{9326394}
Q.~Wu, S.~Zhang, B.~Zheng, C.~You, and R.~Zhang, ``Intelligent reflecting
  surface-aided wireless communications: A tutorial,'' \emph{IEEE Transactions
  on Communications}, vol.~69, no.~5, pp. 3313--3351, 2021.

\bibitem{B15}
X.~Zhao, H.~Liu, S.~Gong, X.~Ju, C.~Xing, and N.~Zhao, ``Dual-functional {MIMO}
  beamforming optimization for {RIS-Aided} integrated sensing and
  communication,'' \emph{IEEE Transactions on Communications}, vol.~72, no.~9,
  pp. 5411--5427, 2024.

\bibitem{B3}
R.~Liu, M.~Li, Q.~Liu, and A.~Lee~Swindlehurst, ``{SNR/CRB}-constrained joint
  beamforming and reflection designs for {RIS-ISAC} systems,'' \emph{IEEE
  Transactions on Wireless Communications}, vol.~23, no.~7, pp. 7456--7470,
  2024.

\bibitem{B8}
Z.~Xing, R.~Wang, and X.~Yuan, ``Joint active and passive beamforming design
  for reconfigurable intelligent surface enabled integrated sensing and
  communication,'' \emph{IEEE Transactions on Communications}, vol.~71, no.~4,
  pp. 2457--2474, 2023.

\bibitem{S1}
Z.~Wang, X.~Mu, and Y.~Liu, ``{STARS} enabled integrated sensing and
  communications,'' \emph{IEEE Transactions on Wireless Communications},
  vol.~22, no.~10, pp. 6750--6765, 2023.

\bibitem{C1}
Z.~Yu, H.~Ren, C.~Pan, G.~Zhou, B.~Wang, M.~Dong, and J.~Wang, ``Active
  {RIS}-aided {ISAC} systems: Beamforming design and performance analysis,''
  \emph{IEEE Transactions on Communications}, vol.~72, no.~3, pp. 1578--1595,
  2024.

\bibitem{B12}
Y.~Xu, Y.~Li, J.~A. Zhang, M.~D. Renzo, and T.~Q.~S. Quek, ``Joint beamforming
  for {RIS}-assisted integrated sensing and communication systems,'' \emph{IEEE
  Transactions on Communications}, vol.~72, no.~4, pp. 2232--2246, 2024.

\bibitem{C2}
Q.~Zhu, M.~Li, R.~Liu, and Q.~Liu, ``Cram{\'e}r-rao bound optimization for
  active {RIS}-empowered {ISAC} systems,'' \emph{IEEE Transactions on Wireless
  Communications}, vol.~23, no.~9, pp. 11\,723--11\,736, 2024.

\bibitem{S5A}
S.~Zhang, W.~Hao, G.~Sun, C.~Huang, Z.~Zhu, X.~Li, and C.~Yuen, ``Joint
  beamforming optimization for active {STAR-RIS}-assisted {ISAC} systems,''
  \emph{IEEE Transactions on Wireless Communications}, vol.~23, no.~11, pp.
  15\,888--15\,902, 2024.

\bibitem{11177504}
X.~Cao, P.~Jiang, G.~Zhu, Y.~He, and M.~Guizani, ``Joint antenna position and
  beamforming optimization for movable antenna enabled secure {IRS-ISAC}
  network,'' \emph{IEEE Transactions on Network Science and Engineering}, pp.
  1--15, 2025.

\bibitem{10254508}
R.~S.~P. Sankar, S.~P. Chepuri, and Y.~C. Eldar, ``Beamforming in integrated
  sensing and communication systems with reconfigurable intelligent surfaces,''
  \emph{IEEE Transactions on Wireless Communications}, vol.~23, no.~5, pp.
  4017--4031, 2024.

\bibitem{B6}
X.~Wang, Z.~Fei, and Q.~Wu, ``Integrated sensing and communication for
  {RIS}-assisted backscatter systems,'' \emph{IEEE Internet of Things Journal},
  vol.~10, no.~15, pp. 13\,716--13\,726, 2023.

\bibitem{10746496}
A.~Abdelaziz~Salem, M.~A. Albreem, K.~A. Alnajjar, S.~Abdallah, and M.~Saad,
  ``Integrated cooperative sensing and communication for {RIS}-enabled
  full-duplex cell-free {MIMO} systems,'' \emph{IEEE Transactions on
  Communications}, vol.~73, no.~6, pp. 3804--3819, 2025.

\bibitem{10366288}
X.~Peng, X.~Hu, X.~Gan, and C.~Zhong, ``Joint location sensing and demodulation
  for {IRS}-assisted {ISAC} mmwave {MIMO} systems,'' \emph{IEEE Transactions on
  Communications}, vol.~72, no.~4, pp. 2470--2484, 2024.

\bibitem{4337601}
R.~M. Mealey, ``A method for calculating error probabilities in a radar
  communication system,'' \emph{IEEE Transactions on Space Electronics and
  Telemetry}, vol.~9, no.~2, pp. 37--42, 1963.

\bibitem{4338141}
S.~Kalenichenko and V.~Mikhailov, ``The joint radar targets detecting and
  communication system,'' in \emph{2006 International Radar Symposium}, 2006,
  pp. 1--4.

\bibitem{5776640}
C.~Sturm and W.~Wiesbeck, ``Waveform design and signal processing aspects for
  fusion of wireless communications and radar sensing,'' \emph{Proceedings of
  the IEEE}, vol.~99, no.~7, pp. 1236--1259, 2011.

\bibitem{6824752}
J.~G. Andrews, S.~Buzzi, W.~Choi, S.~V. Hanly, A.~Lozano, A.~C.~K. Soong, and
  J.~C. Zhang, ``What will 5{G} be?'' \emph{IEEE Journal on Selected Areas in
  Communications}, vol.~32, no.~6, pp. 1065--1082, 2014.

\bibitem{9606831}
Y.~Cui, F.~Liu, X.~Jing, and J.~Mu, ``Integrating sensing and communications
  for ubiquitous {IoT}: Applications, trends, and challenges,'' \emph{IEEE
  Network}, vol.~35, no.~5, pp. 158--167, 2021.

\bibitem{A3}
F.~Dong, F.~Liu, Y.~Cui, W.~Wang, K.~Han, and Z.~Wang, ``Sensing as a service
  in 6{G} perceptive networks: A unified framework for {ISAC} resource
  allocation,'' \emph{IEEE Transactions on Wireless Communications}, vol.~22,
  no.~5, pp. 3522--3536, 2023.

\bibitem{A5}
S.~Li, W.~Yuan, C.~Liu, Z.~Wei, J.~Yuan, B.~Bai, and D.~W.~K. Ng, ``A novel
  {ISAC} transmission framework based on spatially-spread orthogonal time
  frequency space modulation,'' \emph{IEEE Journal on Selected Areas in
  Communications}, vol.~40, no.~6, pp. 1854--1872, 2022.

\bibitem{A9}
X.~Wang, Z.~Fei, J.~A. Zhang, and J.~Xu, ``Partially-connected hybrid
  beamforming design for integrated sensing and communication systems,''
  \emph{IEEE Transactions on Communications}, vol.~70, no.~10, pp. 6648--6660,
  2022.

\bibitem{A8}
D.~Xu, X.~Yu, D.~W.~K. Ng, A.~Schmeink, and R.~Schober, ``Robust and secure
  resource allocation for {ISAC} systems: A novel optimization framework for
  variable-length snapshots,'' \emph{IEEE Transactions on Communications},
  vol.~70, no.~12, pp. 8196--8214, 2022.

\bibitem{A10}
Z.~Liu, S.~Aditya, H.~Li, and B.~Clerckx, ``Joint transmit and receive
  beamforming design in full-duplex integrated sensing and communications,''
  \emph{IEEE Journal on Selected Areas in Communications}, vol.~41, no.~9, pp.
  2907--2919, 2023.

\bibitem{A4}
Z.~He, W.~Xu, H.~Shen, D.~W.~K. Ng, Y.~C. Eldar, and X.~You, ``Full-duplex
  communication for {ISAC}: Joint beamforming and power optimization,''
  \emph{IEEE Journal on Selected Areas in Communications}, vol.~41, no.~9, pp.
  2920--2936, 2023.

\bibitem{A2}
X.~Cheng, D.~Duan, S.~Gao, and L.~Yang, ``Integrated sensing and communications
  {(ISAC)} for vehicular communication networks {(VCN)},'' \emph{IEEE Internet
  of Things Journal}, vol.~9, no.~23, pp. 23\,441--23\,451, 2022.

\bibitem{A6}
C.~Xu, B.~Clerckx, S.~Chen, Y.~Mao, and J.~Zhang, ``Rate-splitting multiple
  access for multi-antenna joint radar and communications,'' \emph{IEEE Journal
  of Selected Topics in Signal Processing}, vol.~15, no.~6, pp. 1332--1347,
  2021.

\bibitem{A7}
Z.~Gao, Z.~Wan, D.~Zheng, S.~Tan, C.~Masouros, D.~W.~K. Ng, and S.~Chen,
  ``Integrated sensing and communication with mmwave massive {MIMO}: A
  compressed sampling perspective,'' \emph{IEEE Transactions on Wireless
  Communications}, vol.~22, no.~3, pp. 1745--1762, 2023.

\bibitem{B2}
J.~Zuo, Y.~Liu, C.~Zhu, Y.~Zou, D.~Zhang, and N.~Al-Dhahir, ``Exploiting {NOMA}
  and {RIS} in integrated sensing and communication,'' \emph{IEEE Transactions
  on Vehicular Technology}, vol.~72, no.~10, pp. 12\,941--12\,955, 2023.

\bibitem{B4}
M.~Luan, B.~Wang, Z.~Chang, T.~H?m?l?inen, and F.~Hu, ``Robust beamforming
  design for {RIS}-aided integrated sensing and communication system,''
  \emph{IEEE Transactions on Intelligent Transportation Systems}, vol.~24,
  no.~6, pp. 6227--6243, 2023.

\bibitem{B5}
Z.~Chen, P.~Chen, Z.~Guo, Y.~Zhang, and X.~Wang, ``A {RIS}-based vehicle {DOA}
  estimation method with integrated sensing and communication system,''
  \emph{IEEE Transactions on Intelligent Transportation Systems}, vol.~25,
  no.~6, pp. 5554--5566, 2024.

\bibitem{B9}
W.~Lyu, S.~Yang, Y.~Xiu, Y.~Li, H.~He, C.~Yuen, and Z.~Zhang, ``{CRB}
  minimization for {RIS}-aided mmwave integrated sensing and communications,''
  \emph{IEEE Internet of Things Journal}, vol.~11, no.~10, pp.
  18\,381--18\,393, 2024.

\bibitem{B13}
K.~Chen, C.~Qi, O.~A. Dobre, and G.~Y. Li, ``Simultaneous beam training and
  target sensing in {ISAC} systems with {RIS},'' \emph{IEEE Transactions on
  Wireless Communications}, vol.~23, no.~4, pp. 2696--2710, 2024.

\bibitem{S6}
N.~Xue, X.~Mu, Y.~Liu, and Y.~Chen, ``{NOMA}-assisted full space
  {STAR-RIS-ISAC},'' \emph{IEEE Transactions on Wireless Communications},
  vol.~23, no.~8, pp. 8954--8968, 2024.

\bibitem{S9}
G.~Sun, Y.~Zhang, W.~Hao, Z.~Zhu, X.~Li, and Z.~Chu, ``Joint beamforming
  optimization for {STAR-RIS} aided {NOMA ISAC} systems,'' \emph{IEEE Wireless
  Communications Letters}, vol.~13, no.~4, pp. 1009--1013, 2024.

\bibitem{S11}
S.~Zhang, W.~Hao, G.~Sun, Z.~Zhu, X.~Li, and Q.~Wu, ``Joint beamforming design
  for the {STAR-RIS}-enabled {ISAC} systems with multiple targets and multiple
  users,'' \emph{IEEE Transactions on Communications}, vol.~73, no.~1, pp.
  693--708, 2025.

\bibitem{M1}
D.~Han, P.~Wang, W.~Ni, W.~Wang, A.~Zheng, D.~Niyato, and N.~Al-Dhahir,
  ``Multi-functional {RIS} integrated sensing and communications for 6{G}
  networks,'' \emph{IEEE Transactions on Wireless Communications}, vol.~24,
  no.~2, pp. 1146--1161, 2025.

\bibitem{BD1}
D.~Wang, Z.~Wang, W.~Yang, H.~Zhao, Y.~He, L.~Li, Z.~Wei, and F.~Zhou,
  ``Enhanced {ISAC} framework for moving target assisted by beyond-diagonal
  {RIS}: Accurate localization and efficient communication,'' \emph{IEEE
  Transactions on Network Science and Engineering}, vol.~12, no.~5, pp.
  4299--4315, 2025.

\bibitem{S13A}
Z.~Yigit and E.~Basar, ``Hybrid {STAR-RIS} enabled integrated sensing and
  communication,'' \emph{IEEE Transactions on Communications}, vol.~73, no.~9,
  pp. 8289--8300, 2025.

\bibitem{C5}
V.~Kumar and M.~Chafii, ``Beamforming design for secure {RIS}-enabled {ISAC}:
  Passive {RIS} versus active {RIS},'' \emph{IEEE Transactions on Wireless
  Communications}, vol.~24, no.~9, pp. 7719--7732, 2025.

\bibitem{S15A}
L.~Guo, J.~Jia, X.~Mu, Y.~Liu, J.~Chen, and X.~Wang, ``Joint secure and covert
  communications for active {STAR-RIS} assisted {ISAC} systems,'' \emph{IEEE
  Transactions on Wireless Communications}, vol.~24, no.~9, pp. 7501--7516,
  2025.

\bibitem{C7}
Z.~Wu and W.~Zhang, ``Joint transmit and reflect beamforming design for
  active-{RIS}-assisted secure {ISAC} systems,'' \emph{IEEE Communications
  Letters}, vol.~29, no.~8, pp. 1769--1773, 2025.

\bibitem{B17}
Z.~Liu, W.~Chen, Q.~Wu, Z.~Li, Q.~Wu, N.~Cheng, and J.~Li, ``Beamforming design
  and multi-user scheduling in transmissive {RIS} enabled distributed
  cooperative {ISAC} networks with {RSMA},'' \emph{IEEE Transactions on
  Communications}, pp. 1--1, 2025.

\bibitem{S18}
Z.~Liu, X.~Li, H.~Ji, H.~Zhang, and V.~C.~M. Leung, ``Toward
  {STAR-RIS}-empowered integrated sensing and communications: Joint active and
  passive beamforming design,'' \emph{IEEE Transactions on Vehicular
  Technology}, vol.~72, no.~12, pp. 15\,991--16\,005, 2023.

\bibitem{BD2}
T.~L. Nguyen, G.~Kaddoum, B.~Selim, and C.~Assi, ``Beyond diagonal {RIS} for
  {ISAC} network: Statistical analysis and network parameter estimation,'' in
  \emph{ICC 2025 - IEEE International Conference on Communications}, 2025, pp.
  2406--2411.

\bibitem{B11}
Y.~Guo, Y.~Liu, Q.~Wu, X.~Li, and Q.~Shi, ``Joint beamforming and power
  allocation for {RIS} aided full-duplex integrated sensing and uplink
  communication system,'' \emph{IEEE Transactions on Wireless Communications},
  vol.~23, no.~5, pp. 4627--4642, 2024.

\bibitem{S4}
W.~Wei, X.~Pang, C.~Xing, N.~Zhao, and D.~Niyato, ``{STAR-RIS} aided secure
  {NOMA} integrated sensing and communication,'' \emph{IEEE Transactions on
  Wireless Communications}, vol.~23, no.~9, pp. 10\,712--10\,725, 2024.

\bibitem{S16}
Y.~Liu, J.~Zhang, Y.~Han, S.~Jin, X.~Li, and Y.~Ni, ``Joint beamforming design
  for {STAR-RIS} enabled {ISAC} system with statistical {CSI},'' \emph{IEEE
  Transactions on Vehicular Technology}, vol.~74, no.~9, pp. 14\,122--14\,137,
  2025.

\bibitem{BD3}
S.~Zheng and S.~Zhang, ``Beyond diagonal intelligent reflecting surface aided
  integrated sensing and communication,'' \emph{IEEE Transactions on Cognitive
  Communications and Networking}, pp. 1--1, 2025.

\bibitem{S2}
Y.~Wang, Z.~Yang, J.~Cui, P.~Xu, G.~Chen, T.~Q.~S. Quek, and R.~Tafazolli,
  ``Optimizing the fairness of {STAR-RIS} and {NOMA} assisted integrated
  sensing and communication systems,'' \emph{IEEE Transactions on Wireless
  Communications}, vol.~23, no.~6, pp. 5895--5907, 2024.

\bibitem{S3}
M.~Li, S.~Zhang, Y.~Ge, Z.~Li, F.~Gao, and P.~Fan, ``{STAR-RIS} aided
  integrated sensing and communication over high mobility scenario,''
  \emph{IEEE Transactions on Communications}, vol.~72, no.~8, pp. 4788--4802,
  2024.

\bibitem{B14}
X.~Yang, Z.~Wei, Y.~Liu, H.~Wu, and Z.~Feng, ``{RIS}-assisted cooperative
  multicell {ISAC} systems: A multi-user and multi-target case,'' \emph{IEEE
  Transactions on Wireless Communications}, vol.~23, no.~8, pp. 8683--8699,
  2024.

\bibitem{C3}
W.~Hao, Y.~Qu, S.~Zhou, F.~Wang, Z.~Lu, and S.~Yang, ``Joint beamforming design
  for hybrid {RIS}-assisted mmwave {ISAC} system relying on hybrid precoding
  structure,'' \emph{IEEE Internet of Things Journal}, vol.~11, no.~18, pp.
  29\,455--29\,469, 2024.

\bibitem{S7}
J.~Zhang, S.~Gong, W.~Lu, C.~Xing, N.~Zhao, D.~W.~K. Ng, and D.~Niyato, ``Joint
  design for {STAR-RIS} aided {ISAC}: Decoupling or learning,'' \emph{IEEE
  Transactions on Wireless Communications}, vol.~23, no.~10, pp.
  14\,365--14\,379, 2024.

\bibitem{S8}
Z.~Liu, X.~Li, H.~Ji, H.~Zhang, and V.~C.~M. Leung, ``Exploiting {STAR-RIS} for
  covert communication in {ISAC} networks under imperfect {CSI},'' \emph{IEEE
  Transactions on Vehicular Technology}, vol.~74, no.~1, pp. 786--802, 2025.

\bibitem{S12A}
Z.~Liu, X.~Li, H.~Ji, and H.~Zhang, ``Active {STAR-RIS}-enabled {ISAC} networks
  against simultaneous eavesdropping and detection attacks,'' \emph{IEEE
  Internet of Things Journal}, vol.~12, no.~11, pp. 16\,841--16\,857, 2025.

\bibitem{S14}
Q.~Zhang, H.~Wu, H.~Li, Z.~Song, and S.~Hou, ``Joint location and beamforming
  design for energy efficient {STAR-RIS}-aided {ISAC} systems,'' \emph{IEEE
  Communications Letters}, vol.~29, no.~1, pp. 140--144, 2025.

\bibitem{C6}
W.~Ma, P.~Zhang, J.~Ye, R.~Guan, X.-P. Li, and L.~Huang, ``Joint antenna
  selection and beamforming design for active {RIS}-aided {ISAC} systems,''
  \emph{IEEE Internet of Things Journal}, vol.~12, no.~14, pp.
  26\,500--26\,513, 2025.

\bibitem{S17A}
S.~Yang, Y.~Jiao, Y.~Wang, F.~Zhao, H.~Guo, Y.~Song, and W.~Hao, ``Robust
  secure energy efficiency optimization for active {STAR-RIS} assisted {ISAC}
  systems,'' \emph{IEEE Open Journal of the Communications Society}, vol.~6,
  pp. 7459--7469, 2025.

\bibitem{B16}
J.~Chen, K.~Wu, J.~Niu, Y.~Li, P.~Xu, and J.~Andrew~Zhang, ``Spectral and
  energy efficient waveform design for {RIS}-assisted {ISAC},'' \emph{IEEE
  Transactions on Communications}, vol.~73, no.~1, pp. 158--172, 2025.

\bibitem{B10}
Y.~Huo, X.~Dong, and N.~Ferdinand, ``Distributed reconfigurable intelligent
  surfaces for energy-efficient indoor terahertz wireless communications,''
  \emph{IEEE Internet of Things Journal}, vol.~10, no.~3, pp. 2728--2742, 2023.

\bibitem{B7}
X.~Qian, X.~Hu, C.~Liu, M.~Peng, and C.~Zhong, ``Sensing-based beamforming
  design for joint performance enhancement of {RIS}-aided {ISAC} systems,''
  \emph{IEEE Transactions on Communications}, vol.~71, no.~11, pp. 6529--6545,
  2023.

\bibitem{B20}
X.~Hu, C.~Liu, M.~Peng, and C.~Zhong, ``{IRS}-based integrated location sensing
  and communication for mmwave {SIMO} systems,'' \emph{IEEE Transactions on
  Wireless Communications}, vol.~22, no.~6, pp. 4132--4145, 2023.

\bibitem{B21}
S.-N. Jin, D.-W. Yue, and H.~H. Nguyen, ``{RIS}-aided cell-free massive {MIMO}
  system: Joint design of transmit beamforming and phase shifts,'' \emph{IEEE
  Systems Journal}, vol.~17, no.~2, pp. 3093--3104, 2023.

\bibitem{B1}
Y.~He, Y.~Cai, H.~Mao, and G.~Yu, ``{RIS}-assisted communication radar
  coexistence: Joint beamforming design and analysis,'' \emph{IEEE Journal on
  Selected Areas in Communications}, vol.~40, no.~7, pp. 2131--2145, 2022.

\bibitem{D1}
M.~I. Ismail, A.~M. Shaheen, M.~M. Fouda, and A.~S. Alwakeel, ``{RIS}-assisted
  integrated sensing and communication systems: Joint reflection and
  beamforming design,'' \emph{IEEE Open Journal of the Communications Society},
  vol.~5, pp. 908--927, 2024.

\bibitem{C4}
M.~Liu, H.~Ren, C.~Pan, B.~Wang, Z.~Yu, R.~Weng, K.~Zhi, and Y.~He, ``Joint
  beamforming design for double active {RIS}-assisted radar-communication
  coexistence systems,'' \emph{IEEE Transactions on Cognitive Communications
  and Networking}, vol.~10, no.~5, pp. 1704--1717, 2024.

\bibitem{S10}
P.~Saikia, A.~Jee, K.~Singh, C.~Pan, W.-J. Huang, and T.~A. Tsiftsis,
  ``{RIS}-aided integrated sensing and communication systems: {STAR-RIS} versus
  passive {RIS}?'' \emph{IEEE Open Journal of the Communications Society},
  vol.~5, pp. 7954--7973, 2024.

\bibitem{B19}
Z.~Yu, X.~Hu, C.~Liu, M.~Peng, and C.~Zhong, ``Location sensing and beamforming
  design for {IRS}-enabled multi-user {ISAC} systems,'' \emph{IEEE Transactions
  on Signal Processing}, vol.~70, pp. 5178--5193, 2022.

\bibitem{10411853}
X.~Yang, Z.~Wei, Y.~Liu, H.~Wu, and Z.~Feng, ``{RIS}-assisted cooperative
  multicell {ISAC} systems: A multi-user and multi-target case,'' \emph{IEEE
  Transactions on Wireless Communications}, vol.~23, no.~8, pp. 8683--8699,
  2024.

\bibitem{9181610}
X.~Hu, C.~Zhong, Y.~Zhang, X.~Chen, and Z.~Zhang, ``Location information aided
  multiple intelligent reflecting surface systems,'' \emph{IEEE Transactions on
  Communications}, vol.~68, no.~12, pp. 7948--7962, 2020.

\bibitem{9724202}
X.~Shao, C.~You, W.~Ma, X.~Chen, and R.~Zhang, ``Target sensing with
  intelligent reflecting surface: Architecture and performance,'' \emph{IEEE
  Journal on Selected Areas in Communications}, vol.~40, no.~7, pp. 2070--2084,
  2022.

\bibitem{5447068}
Z.-q. Luo, W.-k. Ma, A.~M.-c. So, Y.~Ye, and S.~Zhang, ``Semidefinite
  relaxation of quadratic optimization problems,'' \emph{IEEE Signal Processing
  Magazine}, vol.~27, no.~3, pp. 20--34, 2010.

\bibitem{9512486}
R.~Xie, D.~Hu, K.~Luo, and T.~Jiang, ``Performance analysis of joint
  range-velocity estimator with {2D-MUSIC} in {OFDM} radar,'' \emph{IEEE
  Transactions on Signal Processing}, vol.~69, pp. 4787--4800, 2021.

\bibitem{9206044}
W.~Tang, M.~Z. Chen, X.~Chen, J.~Y. Dai, Y.~Han, M.~Di~Renzo, Y.~Zeng, S.~Jin,
  Q.~Cheng, and T.~J. Cui, ``Wireless communications with reconfigurable
  intelligent surface: Path loss modeling and experimental measurement,''
  \emph{IEEE Transactions on Wireless Communications}, vol.~20, no.~1, pp.
  421--439, 2021.

\bibitem{9246254}
H.~Xie, J.~Xu, and Y.-F. Liu, ``Max-min fairness in {IRS}-aided multi-cell
  {MISO} systems with joint transmit and reflective beamforming,'' \emph{IEEE
  Transactions on Wireless Communications}, vol.~20, no.~2, pp. 1379--1393,
  2021.

\end{thebibliography}

\end{document}